\documentclass[a4paper,11pt]{article}
\pdfoutput=1
\usepackage{jcappub}
\usepackage{amsfonts,amsmath,amssymb,amsthm,bbm,hyperref}
\usepackage{graphicx}
\usepackage{xcolor}
\usepackage[T1]{fontenc}
\usepackage{lmodern}
\usepackage[normalem]{ulem}
\usepackage{mathrsfs}
\usepackage{verbatim}

\makeatletter

\def\D{\mathrm{d}}
\def\I{\mathrm{i}}
\def\E{\mathrm{e}}
\def\be{\begin{equation}}
\def\ee{\end{equation}}

\def\del#1#2{\frac{\partial #1}{\partial #2}}

\newcommand{\cH}{{\mathcal H}}

\newcommand{\MP}{M_\mathrm{P}}

\newcommand{\beq}{\begin{eqnarray}}
\newcommand{\eeq}{\end{eqnarray}}
\newcommand{\nn}{\nonumber}

\newcommand{\stkout}[1]{\ifmmode\text{\sout{\ensuremath{#1}}}\else\sout{#1}\fi}

\newcommand{\HdS}{H_\mathrm{dS}}

\newcommand{\ka}{k_\mathrm{a}}
\newcommand{\kb}{k_\mathrm{b}}
\newcommand{\kc}{k_\mathrm{c}}
\newcommand{\kmin}{k_\mathrm{min}}

\newcommand{\Nuni}{\mathcal{N}}

\makeatother

\title{The interacting multiverse \\ and its effect on the \\ cosmic microwave background}

\author[a,b]{Mariam Bouhmadi-L\'{o}pez,}
\author[c,d]{Manuel Kr{\"a}mer,}
\author[a]{Jo\~{a}o Morais}
\author[e,f]{and Salvador Robles-P\'{e}rez}

\affiliation[a]{Department of Theoretical Physics, University of the Basque Country UPV/EHU, \\ P.O.~Box 644, 48080 Bilbao, Spain}
\affiliation[b]{IKERBASQUE, Basque Foundation for Science, \\ 48011 Bilbao, Spain}
\affiliation[c]{Institute of Physics, University of Szczecin, \\ Wielkopolska 15, 70-451 Szczecin, Poland}
\affiliation[d]{Institute for Theoretical Physics, KU Leuven, \\ Celestijnenlaan 200D, 3001 Leuven, Belgium}
\affiliation[e]{Departamento de Matem\'{a}ticas, IES Miguel Delibes, \\ Miguel Hern\'{a}ndez 2, 28991 Torrej\'{o}n de la Calzada, Spain}
\affiliation[f]{Estaci\'{o}n Ecol\'{o}gica de Biocosmolog\'{\i}a, \\ Pedro de Alvarado 14, 06411 Medell\'{\i}n, Spain}

\emailAdd{mariam.bouhmadi@ehu.eus}
\emailAdd{manuel.kraemer@kuleuven.be}
\emailAdd{jviegas001@ikasle.ehu.eus}
\emailAdd{salvador.robles@educa.madrid.org}

\abstract{We study a toy model of a multiverse consisting of canonically quantized universes that interact with each other on a quantum level based on a field-theoretical formulation of the Wheeler--DeWitt equation. This interaction leads to the appearance of a pre-inflationary phase in the evolution of the individual universes. We analyze scalar perturbations within the model and calculate the influence of the pre-inflationary phase onto the power spectrum of these perturbations. The result is that there is a suppression of power on large scales, which can describe well the Planck 2018 data for the cosmic microwave background anisotropies and could thus indicate a possible solution to the observed quadrupole discrepancy.}


\arxivnumber{1809.09133}

\begin{document}
\maketitle

%
%

\section{Introduction}

The idea that our universe is part of a multiverse, which consists of a multitude of other universes that might even have different properties like different physical constants and laws, is intriguing, but will remain speculative as long as one cannot find a way that other parts of the multiverse can have an observable effect onto our universe.

The multiverse hypothesis also comes in multiple flavors (for an overview of the wide variety of forms, see e.g.~\cite{Everett1957, DeWitt1973, Linde1983, Linde1986, Smolin1997, Steinhard2002, Susskind2003, Tegmark2003, Freivogel2004, Tegmark2007, Carr2007, Mersini2008a, Mersini2008b, RP2010, Alonso2012, RP2016,Bouhmadi2007}) and one thus needs to be careful to clearly define what one means when discussing issues about the multiverse.

However, in spite of this variety, two main paradigms can be distinguished. The first could be named the \emph{statistical paradigm} of the multiverse (see \cite{Garriga2013, Vilenkin2014} and references therein), in which the universes do not directly interact with each other and thus only account for a statistical measure of the different values that a physical parameter can take in a single realization of the universe. Predictions in this type of multiverse are ultimately based on Vilenkin's principle of mediocrity \cite{Vilenkin2011}, which somehow extends the Copernican principle and states that our universe is expected to be one of the most probable universes of the multiverse.

The other paradigm of the multiverse can be called the \emph{interacting multiverse} \cite{Bertolami2008, Mersini2008c, Mersini2008d, RP2012, Alonso2012, RP2016}, where quantum interactions and other non-local correlations may exist between the quantum states of two or more universes. These non-local correlations do not rule out the notion of causal closure associated to the universes because that closure is always related to a local definition of causality (which is the only way in which causality is defined in physics).

Moreover, these two ideas of the multiverse are not mutually exclusive because in quantum cosmology a non-interacting multiverse in a given representation turns out to be an interacting one in another quantum representation \cite{Alonso2012, RP2016,Vedral}, so they are complementary.  However, predictions in the interacting multiverse are not restricted to those of a statistical nature because a particular type of interaction between two universes of the multiverse may leave a distinguishable imprint in the properties of a universe like ours \cite{Mersini2008c, Mersini2008d, RP2016, Mersini2017, DiValentino2017a, DiValentino2017b, RP2017c}.

A suitable way to describe a multiverse in a quantum-cosmological setting based on a canonical quantization with the Wheeler--DeWitt equation is the so-called third quantization formalism \cite{Caderni1984, Rubakov1988, McGuigan1988, Strominger1990, RP2009, RP2010, Faizal2012, Gielen2012, Calcagni2012, RP2013, Kim2012, Garay2014}. It consists of considering the wave function of the universe as a field that propagates in the abstract superspace of spacetime geometries and matter fields. Then, the hyperbolic nature of the Wheeler--DeWitt equation \cite{McGuigan1988} ensures that it can be interpreted as the equation to determine the wave function of the universe. Following a parallel procedure to that of a quantum field theory, the wave function of the universe can then be promoted to a quantum operator whose evolution can be decomposed in partial waves with creation and annihilation operators that, in some appropriate representation, might represent the creation and annihilation of universes. 
This cannot always be done in the general superspace of geometries and matter fields, because we need a time-like variable to define the operators for the creation and annihilation of universes \cite{Strominger1990,clausbook,Kuchar1992}. This is similar to the case of a general curved space-time, where a non-well-defined notion of time makes it impossible to give a proper definition of a particle. However, for homogeneous and isotropic space-times the scale factor can formally be seen as the time-like variable of the minisuperspace, which can also be inferred from the Lorentzian signature of the minisupermetric of the minisuperspace \cite{Strominger1990, clausbook, RP2010, Garay2014}. In those cases the formalism simplifies and a quantum field procedure can be applied to the minisuperspace much in the same way as it is done in a usual quantum field theory. 

It has already been shown that applying the third quantization procedure to the model of an eternally inflating universe leads to the picture of a multiverse whose sub-universes exhibit a distinct pre-inflationary phase that appears in the Friedmann equation as an $a^{-6}$-term. In a previous article \cite{Morais2017}, we have analyzed a toy model of a universe with scalar perturbations undergoing an evolution like one of those sub-universes and found that the pre-inflationary phase leads to a suppression of power on the largest scales followed by a bump resulting in an enhancement {of power}. Given that the most recent data from measurements of the cosmic microwave background (CMB) \cite{Akrami:2018vks} are suggesting that there is a suppression of power on large scales, often called CMB quadrupole anomaly, we found that in order to get a sizeable effect for the suppression from the pre-inflationary phase to explain the quadrupole anomaly in the CMB, the bump is enhanced too much to be compatible with the CMB data.

The model described above did not yet include an explicit interaction between the different universes, which will be the purpose of the present article. In previous work \cite{Alonso2012}, such an interaction led to a different pre-inflationary phase, and we shall see that the modification of the CMB anisotropy spectrum arising from such a pre-inflationary phase coincides better with the CMB data and in principle could explain better the quadrupole issue as discussed in our previous work \cite{Morais2017}. However, it turns out that a full third quantization is not needed to describe an interacting multiverse as long as the interaction is weak enough not to involve the creation and annihilation of universes. Following the analogy with particle physics, we are not considering here strong interactions that would lead to the creation and annihilation of particles, but weak interactions that slightly modify the energy (frequency) of the given particles. Hence, a field-theoretical formulation of the Wheeler--DeWitt equation inspired by the third quantization formalism is sufficient.

Of course, there have been other approaches to explain the observed quadrupole anomaly going from a proposed fast-roll inflationary phase before standard inflation \cite{Contaldi:2003zv,Boyanovsky:2006pm}, bouncing as well as cyclic universes \cite{Piao:2003zm,Piao:2005ag,Liu:2013kea}, a pre-inflationary era dominated by radiation \cite{Powell:2006yg,Wang:2007ws} to a matter-dominated era before inflation that is caused by remnants of primordial micro black holes \cite{Scardigli:2010gm}. More recent proposals include a topological defect phase before slow-roll inflation \cite{BouhmadiLopez:2012by} and effects of a compactification mechanism before inflation \cite{Kontou:2017xhp}.

This article is organized as follows. In Section 2, we present our interacting universe model inspired by the third quantization formalism and show how to use a semiclassical approximation to derive the evolution of the individual universes and their perturbations. This is followed by Section 3, which deals with the numerical calculation of the scalar power spectra and the corresponding CMB anisotropy spectra for the specific models we chose to discuss in detail. Finally, we give our conclusions in Section 4. We include as well an appendix to give details of a semiclassical approximation we use.


\section{Modelling an interacting multiverse}
\label{section2}

Let us consider a flat Friedmann--Lema\^itre--Robertson--Walker universe with scale factor $a$ that contains a scalar field $\varphi$ with potential $\mathcal{V}(\varphi)$ as well as perturbations of the metric and the scalar field. The scalar metric perturbations can be written using four scalars $A$, $B$, $\Xi$, $E$ and the conformal time $\eta$ in the following way \cite{Bassett:2005xm}
\begin{align}
	\D s^2 = 
	-a^2\left(1+2A\right) \D\eta^2 
	+ 2a^2\partial_iB \,\D x^i \D\eta
	+a^2\left[\left(1-2\,\Xi\right)\delta_{ij} + 2\partial_i\partial_j E\right]\D x^i \D x^j
	\,.
\end{align}
and together with the scalar field perturbations $\delta\varphi$, they can be combined into the gauge-invariant Mukhanov--Sasaki variable \cite{Mukhanov:1990me}
\be
v = a\left[\delta\varphi + \left(\frac{\varphi'}{\mathscr{H}}\right)\Xi\right],
\ee
where the following definitions were used: $(\cdot)':=\D(\cdot)/\D\eta$ and $\mathscr{H}:=a'/a$.

Denoting the Fourier transform of $v$ with respect to the wave number $k:=|\vec{k}|$ as $v_k$, we can write down the action describing our perturbed universe model:
\begin{align} \label{genaction}
S = \frac{1}{2}\int
\D\eta\,\left\lbrace-\,\frac{3}{4\pi G}\,(a')^2 +
  a^2\,(\phi')^2-2\,a^4\,\mathcal{V}(\phi)
  +\sum_{k}
\,\Bigl[ v_{k}^\prime{v_{k}^*}^\prime
+\omega^2_{k}\,v_{k}v_{k}^*\Bigr] \right\rbrace.
\end{align}
where $G$ is the gravitational constant and $\omega^2_k(\eta)$ is given using $z := a (\varphi'/\mathscr{H})$ by
\be
\omega^2_k(\eta) := k^2 - \frac{z''}{z}\,.
\ee
We have absorbed the length scale that one needs to introduce into the action $\eqref{genaction}$ due to integrating over a finite volume and due to the discretization of the modes $k$ into the quantities $a$, $\eta$, $k$ and $v$ as shown in \cite{bkk15,bkk16}.

We can also express the potential $z''/z$ appearing as an effective mass term in the above equation in terms of the slow-roll parameters $\epsilon$, $\delta$ and $\xi$, which are defined using the Hubble parameter $H:=\dot{a}/a$ with respect to the cosmic time derivative $(\cdot)\dot{}:=\D(\cdot)/\D t$ as \cite{Bassett:2005xm}
\begin{align}
	\epsilon
	:=
	-\frac{\dot{H}}{H^2} 
	= 
	4\pi\frac{\dot{\varphi}^2}{H^2}
	\,,
	\qquad
	\delta
	:=
	\epsilon
	- \frac{\dot{\epsilon}}{2H\epsilon}
	=
	-\frac{\ddot{\varphi}}{H\dot{\varphi}}
	\,,
	\qquad
	\textrm{and}
	\qquad
	\xi 
	:=
	2\epsilon\left(\epsilon+\delta\right) - \frac{\dot{\epsilon}+\dot{\delta}}{H}
	\,,
\end{align}
and thus we end up with the expression \cite{Bassett:2005xm}
\begin{align} \label{zzquot}
	\frac{z''}{z}
	=
	\left(aH\right)^2\left[
		2 
		+ 2\epsilon 
		- 3\delta 
		+3\epsilon^2
		-5\epsilon\delta
		+\delta^2
		+\xi^2
	\right].
\end{align}
Canonically quantizing our model and using the Laplace--Beltrami factor ordering, we arrive at the following Wheeler--DeWitt equation \cite{bkk15,bkk16}
\begin{align}
	\Bigg[
		\frac{2\pi\hbar^2 G}{3}\left(\del{^2}{a^2} 
		+ \frac{1}{a}\,\del{}{a}\right) 
		- \frac{\hbar^2}{2a^2}\,\del{^2}{\varphi^2}
		+ a^4\,\mathcal{V}(\varphi) &
	\nn\\
		+ \sum_k \left(-\,\frac{\hbar^2}{2}\,\del{^2}{v_k^2} 
		+ \frac{\omega^2_k(a)}{2}\,v_k^2\right)
	&\Bigg]\Psi\big(a,\varphi,\{v_k\}\big)= 0\,.
\label{WDW_raw}
\end{align}
Note that, in principle, the ratio $z''/z$ contains canonical momenta with regard to the background quantities $a$ and $\varphi$ that in a full quantization would need to be replaced by derivatives (cf.~\cite{langlois1994}). However, we assume in the following that our background is quasi-de Sitter, such that we can employ the slow-roll regime, and as such use expression \eqref{zzquot} keeping the slow-roll parameters as classical variables and replacing $H$ by the quasi-static Hubble parameter instead of using the expressions given e.g.~in \cite{langlois1994}.
 
In order to simplify the notation, we set $\hbar \equiv 1$ and introduce a rescaled Planck mass $m_\text{P}$ defined as
\begin{align}
	m_\text{P}^2 
	:= 
	\frac{3}{4\pi G}
	\,.
\end{align}
As mentioned above, we assume that our background is quasi-de Sitter and thus we demand that the scalar field is approximately constant and we can therefore neglect the $\varphi$-kinetic term and set the potential to be
\begin{align}
	 \label{Vmp}
	\mathcal{V}(\varphi) 
	= \frac{3}{8\pi G}\,\HdS^2 
	= \frac{1}{2}\,m_\text{P}^2\,\HdS^2
	\,,
\end{align}
where $\HdS$ is the quasi-constant Hubble parameter arising from the quasi-de Sitter evolution of this model universe caused by the constant scalar-field potential. Thus we end up with the following Wheeler--DeWitt equation
\begin{align}
	\Biggl[
		\frac{1}{m_\text{P}^2}\left(\del{^2}{a^2} 
		+ \frac{1}{a}\,\del{}{a}\right)
		+ m_\text{P}^2\,a^4\HdS^2 
		+ \sum_k\left(-\,\del{^2}{v_k^2} 
		+ \omega^2_k(a)\,v_k^2\right)
	\Biggr]\Psi\big(a,\{v_k\}\big)
	= 0
	\,. 
	\label{WDW_allk}
\end{align}
Our aim is now to consider a multiverse model in the spirit of the third quantization formalism. This would in principle consist in promoting the wave function $\Psi$ to an operator. However, also by means of a field-theoretical formulation of the Wheeler--DeWitt equation, we can eventually introduce an interaction scheme at the quantum level.

We thus consider a ``super-action'' $S_{\text{3Q}}$ defined on the symmetry-reduced superspace $(a,\{v_k\})$
\begin{align}
	\mathcal{S}_\text{3Q} 
	= \int\D a\prod_k\D v_k\,a&\left\{
		 -\,\frac{1}{m_\text{P}^2} \del{\Psi^*}{a}\,\del{\Psi}{a} 
		 + m_\text{P}^2\,a^4\HdS^2\,\Psi^*\Psi 
		 + \sum_k \left[\del{\Psi^*}{v_k}\,\del{\Psi}{v_k} + \omega^2_k(a)\,v_k^2\,\Psi^*\Psi\right]
	\right\},
\end{align}
from which the Wheeler--DeWitt equation \eqref{WDW_allk} can be recovered by means of the Euler--Lagrange equation
\be
\del{}{a}\!\left(\frac{\delta \mathcal{L}}{\delta(\partial_a\Psi^*)}\right) + \sum_k\del{}{v_k}\!\left(\frac{\delta \mathcal{L}}{\delta(\partial_{v_k}\Psi^*)}\right) - \frac{\delta \mathcal{L}}{\delta \Psi^*} = 0\,.
\ee
Using a Legendre transform, the corresponding super-Hamiltonian is then given by
\be
\mathcal{H}_\text{3Q} = -\,a\left[ P_{\Psi^*}P_{\Psi} + \sum_k \del{\Psi^*}{v_k}\,\del{\Psi}{v_k} + \left(m_\text{P}^2\,a^4\HdS^2 + \sum_k \omega^2_k(a)\,v_k\right)\Psi^*\Psi\right],
\label{Ham_single}
\ee
where the conjugate momentum $P_{\Psi}$ reads
\be
P_{\Psi} = \frac{\delta \mathcal{L}_\text{3Q}}{\delta(\partial_a\Psi^*)} = -\,\frac{a}{m_\text{P}^2}\,\del{\Psi}{a}\,.
\ee
A full third quantization in the original sense would now correspond to promoting the wave function $\Psi$ to an operator or, alternatively, to define a higher-order wave functional $\widetilde{\Psi}_\text{3Q}[\Psi(a,v_k)]$ that fulfills a Schr\"odinger-like functional equation with the scale factor taking the role of time. However, we do not take this route and instead only use the super-Hamiltonian \eqref{Ham_single} to construct an interaction scheme between quantum universes.

For this purpose, we assume that we are dealing with a multiverse consisting of $\mathcal{N}$ universes, where each of these universes with label $J = 1,\ldots, \Nuni$ is described by a wave function $\Psi_J$ that obeys a Wheeler--DeWitt equation analogous to \eqref{WDW_allk},
\begin{align}
	\Biggl[
		\frac{1}{m_\text{P}^2}\left(\del{^2}{a^2} 
		+ \frac{1}{a}\,\del{}{a}\right)
		+ m_\text{P}^2\,a^4\HdS^2 
		+ \sum_k\left(-\,\del{^2}{v_{k}^2} 
		+ \omega^2_{k,J}(a)\,v_{k}^2\right)
	\Biggr]\Psi_J\big(a,\{v_{k}\}\big)
	= 0 \,.
	\label{WDWJ_allk}
\end{align}
Here, we have not indicated that the variables $v_k$ differ in the individual universes and thus depend on the label $J$. However, as will be clear later on after performing a semiclassical approximation to obtain the effective evolution of the individual universes, our quantization scheme does in fact lead to the result that the perturbations evolve differently in the individual universes. In order not to obfuscate the notation too much, we have refrained from indicating this dependence on the label $J$ except for the quantity $\omega_{k,J}(a)$, which -- as commented above -- remains unquantized.

The full super-Hamiltonian of this multiverse, $\cH_\text{multi}$, is made up by the sum of the individual super-Hamiltonians $\cH^J_\text{3Q}$ given by an expression corresponding to \eqref{Ham_single},
\be
\mathcal{H}^J_\text{3Q} = -\,a\left[ P_{\Psi^*_J}P_{\Psi_J} + \sum_k \del{\Psi^*_J}{v_{k}}\,\del{\Psi_J}{v_{k}} + \left(m_\text{P}^2\,a^4\HdS^2 + \sum_k \omega^2_{k,J}(a)\,v_{k}\right)\Psi^*_J\Psi_J\right],
\label{Ham_J_3Q}
\ee
plus an interaction term between different universes, i.e.~different wave functions $\Psi_J$. For this interaction, we assume one of the simplest possible models, which is a nearest-neighbor interaction analogous to a closed chain of harmonic oscillators coupled by springs, like the one described in  \cite{Lievens2006,Lievens2007,RP2016}.
This interaction term is chosen essentially ad hoc. The choice arises from the motivation to construct a model that can be treated analytically and leads to potentially observable effects. We restricted ourselves to an interaction of quadratic order, because higher-order interactions would require the use of a full quantum-field-theoretical framework, i.\,e.~full third quantization, instead of the field-theoretical scheme that we present here.
With such an interaction, the super-Hamiltonian of the multiverse can be expressed as follows
\be\label{totalH}
\mathcal{H}_\text{multi} = \sum_J \Bigl[ \mathcal{H}^J_\text{3Q} - a\,m_\text{P}^2\,\bar{\lambda}^2(a) \left(\Psi_{J+1} - \Psi_J\right)^*\left(\Psi_{J+1} - \Psi_J\right) \Bigr] ,
\ee
where we also assumed that $\Psi_{\Nuni+J} = \Psi_J$. Furthermore, we have included a coupling function $\bar{\lambda}(a)$, for which we assume phenomenologically that it can depend on $a$. The exact form of this function has presently to be chosen ad hoc. Eventually, the hope is that a yet to be found more fundamental theory of quantum gravity can further justify the form of such a function. For now, the guiding principle we take is based on the analysis presented in \cite{RP2016}, which restricts the exponent in $\bar{\lambda}(a) \propto a^n$ to fulfill the condition $n \leq 2$, such that the vacuum decay described in \cite{RP2016} cannot grow with the scale factor. This also translates to the requirement that the interaction should decay with the growth of the scale factor, such that the quantum effects are stronger in the past.

We can diagonalize the multiverse Hamiltonian \eqref{totalH} by applying the discrete Fourier transformation
\be \label{Fouriertransf}
	\Psi_J = \frac{1}{\sqrt{\Nuni}} \sum_{\tilde{K}}  \exp\!\left(-\,\frac{2 \pi \I\,\tilde{K} J}{\Nuni}\right) \Psi_{\tilde{K}} 
	\,, 
	\qquad
	P_{\Psi_J} = \frac{1}{\sqrt{\Nuni}} \sum_{\tilde{K}} \exp\!\left(\frac{2 \pi \I\, \tilde{K} J}{\Nuni}\right) P_{\Psi_{\tilde{K}}} 
	\,.
\ee
That way we get for the interaction term in \eqref{totalH}, which we define as $\Phi_\text{int}$:
\begin{align}
	\Phi_\text{int} 
	:=&~ \sum_J\left(\Psi_{J+1} - \Psi_J\right)\left(\Psi_{J+1} - \Psi_J\right)^* 
	= 2\sum_J\Psi_J\Psi_J^* -\sum_J\left(\Psi_{J+1}\Psi_J^*+\Psi_{J}\Psi_{J+1}^*\right) \nonumber\\
	=&~ \frac{2}{\Nuni}\sum_J\sum_{\tilde{K},\tilde{K}'}\E^{\frac{2\pi\I}{\Nuni}(\tilde{K}'-\tilde{K})J}\Psi_{\tilde{K}}\Psi_{\tilde{K}'}^* 
	\nn\\
	&~-\frac{1}{\Nuni}\sum_J\sum_{\tilde{K},\tilde{K}'}\left(\E^{\frac{2\pi\I}{\Nuni}\left[(\tilde{K}'-\tilde{K})J-\tilde{K}\right]}\Psi_{\tilde{K}}\Psi_{\tilde{K}'}^*+\E^{\frac{2\pi\I}{\Nuni}\left[(\tilde{K}'-\tilde{K})J+\tilde{K}'\right]}\Psi_{\tilde{K}}\Psi_{\tilde{K}'}^*\right).
\end{align}
Using the relation $\sum_{J=1}^{\Nuni} \exp\left[\frac{2\pi\I}{\Nuni}(\tilde{K}'-\tilde{K})J\right] = \Nuni\,\delta_{\tilde{K}\tilde{K}'}$, we can simplify the interaction term to
\begin{align}
	\Phi_\text{int}
	&= 
	2\sum_{\tilde{K}} \Psi_{\tilde{K}}\Psi_{\tilde{K}}^* 
	- \sum_{\tilde{K}}\left(
		\E^{-\frac{2\pi\I}{\Nuni}\tilde{K}}\Psi_{\tilde{K}}\Psi_{\tilde{K}}^*
		+ \E^{\frac{2\pi\I}{\Nuni}\tilde{K}}\Psi_{\tilde{K}}\Psi_{\tilde{K}}^*
	\right) 
	\nonumber\\
	&= 
	\sum_{\tilde{K}}\left[2-2\cos\left(\frac{2\pi\tilde{K}}{\Nuni}\right)\right]\Psi_{\tilde{K}}\Psi_{\tilde{K}}^* 
	= \sum_{\tilde{K}}4\sin^2\left(\frac{\pi\tilde{K}}{\Nuni}\right)\Psi_{\tilde{K}}\Psi_{\tilde{K}}^*
	\,.
\end{align}
We assume that we deal with a large number $\Nuni$ of universes, such that we can write
\begin{align}
	\Phi_\text{int} 
	= 
	\sum_{\tilde{K}}\frac{4\pi^2}{\Nuni^2}\,
	\tilde{K}^2\,\Psi_{\tilde{K}}\Psi_{\tilde{K}}^*\,.
\end{align}
The diagonalized multiverse super-Hamiltonian then reads:
\be
	\mathcal{H}_\text{multi} 
	= 
	\sum_{\tilde{K}}\left(  \mathcal{H}^{\tilde{K}}_\text{3Q}-\,\frac{a}{2}\,m_\text{P}^2\,\lambda^2(a)\,
	\tilde{K}^2\,\Psi_{\tilde{K}}\Psi_{\tilde{K}}^*\right),
	\label{Ham_multi_diag}
\ee
where we used the definitions
\be
	\lambda(a) 
	:= 
	\frac{2\pi\bar\lambda(a)}{\Nuni}
\ee
and
\be
	\mathcal{H}^{\tilde{K}}_\text{3Q} 
	:= 
	-\,\frac{a}{2}\left[ 
		P_{\Psi_{\tilde{K}}}^2 
		+ \sum_k \left(\del{\Psi_{\tilde{K}}}{v_{k}}\right)^2 
		+ \left(m_\text{P}^2\,a^4\HdS^2 
		+ \sum_k \omega^2_{k,\tilde{K}}(a)\,v_{k}\right)\Psi_{\tilde{K}}\Psi_{\tilde{K}}^*
	\right].
	\label{Ham_singleK}
\ee
Here again, we have only indicated a dependence on the label $\tilde{K}$ for the quantity $\omega_{k,\tilde{K}}(a)$.
We now go back to considering only a single universe described by the wave function $\Psi_{\tilde{K}}(a,\{v_{k}\})$ within this interaction scheme. From the super-Hamiltonian \eqref{Ham_multi_diag} specified to a single universe with label $\tilde{K}$, 
\be
\mathcal{H}^{\tilde{K}}_\text{multi} = \mathcal{H}^{\tilde{K}}_\text{3Q}-\,\frac{a}{2}\,m_\text{P}^2\,\lambda^2(a)\,\tilde{K}^2\,\Psi_{\tilde{K}}\Psi_{\tilde{K}}^*\,,
\ee
we obtain the Wheeler--DeWitt equation
\be
\Biggl[\frac{\E^{-2\alpha}}{m_\text{P}^2}\,\del{^2}{\alpha^2}+m_\text{P}^2\left(\E^{4\alpha}\HdS^2 + \lambda(\alpha)^2 \tilde{K}^2\right) + \sum_k\left(-\del{^2}{v_{k}^2} + \omega^2_{k,\tilde{K}}(\alpha)\,v_{k}^2\right) \Biggr]\Psi_{\tilde{K}}(\alpha,\{v_{k}\})
= 0\,,
\label{WDW_Ktilde}
\ee
where we introduced for simplification the logarithmic scale factor $\alpha$ defined with respect to a reference scale factor $a_0$ as
\be
\alpha := \ln\!\left(\frac{a}{a_0}\right).
\ee
Let us now compactify the notation further by introducing the potential
\be
\mathcal{V}_{\tilde{K}}(\alpha) := \E^{4\alpha}\HdS^2 + \lambda(\alpha)^2 \tilde{K}^2
\,,
\ee
as well as the Hamiltonian for the perturbations
\be
\mathcal{H}_{k,\tilde{K}} = \frac{1}{2} \left(-\,\del{^2}{v_{k}^2} + \omega^2_{k,\tilde{K}}(\eta)\,v_{k}^2\right).
\ee
The Wheeler--DeWitt equation \eqref{WDW_Ktilde} thus reads
\be
\Biggl[\frac{\E^{-2\alpha}}{m_\text{P}^2}\,\del{^2}{\alpha^2}+m_\text{P}^2\mathcal{V}_{\tilde{K}}
(\alpha) + 2\sum_k\mathcal{H}_{k,\tilde{K}} \Biggr]\Psi_{\tilde{K}}
(\alpha,\{v_{k}\})
= 0\,.
\label{WDW_Ktilde2}
\ee
We now use the semiclassical approximation presented in \cite{hh85,ck87} in order to derive the Hamilton--Jacobi equation for the background as well as the Schr\"odinger equation for the perturbations, which both arise from this Wheeler--DeWitt equation. We show this derivation in detail in the appendix \ref{app_semi}.

The approximation is based on writing the wave function of one of the individual universes as
\be
\Psi_{\tilde{K}}\big(\alpha,\{v_{k}\}\big) = \E^{\I\,m_\text{P}^2\,S_{0,\tilde{K}}}\prod_{k}\psi_{k,\tilde{K}}\big(\alpha,v_{k}\big)
\ee
and with this we arrive at the following Hamilton--Jacobi equation for the background
\be \label{HJeq1}
\left(\frac{\partial S_{0,\tilde{K}}}{\partial\alpha}\right)^2 
- \E^{6\alpha}H_0^2 -  \E^{2\alpha}\lambda(a)^2 \tilde{K}^2 =0\,,
\ee
as well as a Schr\"odinger equation for the perturbation modes:
\be \label{Schr_vk1}
\I\,\del{}{\eta}\,\psi_{k,\tilde{K}} = \frac{1}{2}\left(-\,\del{^2}{v_{k}^2} + \omega^2_{k,\tilde{K}}(\eta)\,v_{k}^2\right)\psi_{k,\tilde{K}}\,.
\ee
Here, we would like to point out that in the individual universes with distinct values of the $\tilde{K}$, the semi-classical equations \eqref{HJeq1} and \eqref{Schr_vk1} that describe the evolution of the cosmological background and of the linear perturbations will have different solutions. As such, strictly speaking both the semi-classical solutions of the scale factor $a$ and the Mukhanov--Sasaki variable $v_k$ are functions of the parameter $\tilde{K}$, as is the wave function $\psi_{k,\tilde{K}}$. As mentioned above, in order not to burden the notation, we have refrained from explicitly using a $\tilde{K}$-index for these variables. However, in the following subsections, where we derive solutions with an explicit dependence on $\tilde{K}$, we will indicate this $\tilde{K}$-dependence with an index.

\subsection{The evolution of the individual universes in the interacting multiverse}

The Hamilton--Jacobi equation \eqref{HJeq1} describing the evolution of the individual universes can be equivalently written in terms of the scale factor $a$ as
\be \label{HJ_sq}
\frac{\partial S_{0,\tilde{K}}}{\partial a} = \pm\sqrt{a^4\,H_0^2 + \lambda^2(a) \tilde{K}^2}\,.
\ee
By using the classical relation of the canonical momentum $p_a$ of $a$, we obtain
\be
\frac{\partial S_{0,\tilde{K}}}{\partial a} = p_a = -\,a\, \frac{\D a }{\D t}
\,,
\ee
and by choosing the minus sign in Eq.~\eqref{HJ_sq}, which corresponds to considering an expanding universe, we end up with the following Friedmann equation for the background of a single universe within the interacting multiverse we are considering:
\be\label{Fried_bg}
H^2_{\tilde{K}}(a) \equiv \left( \frac{1}{a} \frac{\D a }{\D t} \right)^2 = \HdS^2 + \frac{\lambda^2(a) \tilde{K}^2}{a^4} \,.
\ee
It can be clearly seen that our constructed interaction scheme between the universes making up the multiverse affects the effective form of the Friedmann equation and it appears as a non-local correction of the original Friedmann equation. The value of the mode $\tilde{K}$, or rather its product with the prefactor in $\lambda(a)^2$, is taken to be small given that we are considering here small interactions that make the evolution of the universe slightly depart from the non-interacting case. Of course, as the scale factor approaches the Planck length the approximation breaks down and we would enter a primordial era with large interactions giving rise to the creation and annihilation of universes (like in particle physics). However, in spite of the smallness of the constants in the correction term, it may induce a measurable departure from the evolution of a non-interacting universe.

In \cite{RP2012}, where this interaction scheme was also used to discuss the effect of this scheme on vacuum decay, the boundary condition that assumes that the vacuum decay must be exponentially suppressed at large values of the scale factor imposes the restriction $\lambda^2(a) \leq a^4$. However, even with this restriction, there is still plenty of room for distinguishable effects of the interactions in the properties of a single universe like ours. Eventually, as mentioned above, the form of the coupling function, $\lambda(a)$, should be derived from some effective low-energy limit of the underlying theory of quantum gravity.

In this paper we shall restrict our attention to the particular cases:
\begin{itemize}
\item[i)] $\lambda(a) = \lambda_*$, where $\lambda_*$ is a constant, and
\item[ii)] $\lambda(a) \propto \sqrt{a}$.
\end{itemize}
The former has an effect on the evolution of the universe that is equivalent to radiation $(H^2(a) \propto a^{-4}$), and the latter has an effect that can be regarded as the one caused by matter-like content in the early universe ($H^2(a) \propto a^{-3}$).

We thus write $\lambda(a)$ in terms of a power-law dependence with respect to the scale factor, $\lambda(a)=\lambda_*(a/a_*)^n$, with a constant $\lambda_*$ that has the dimension of length, an arbitrary reference scale factor $a_*$ and an exponent $n$ that in our case takes the values $n=0$ and $n=1/2$. The Friedmann equation \eqref{Fried_bg} can then be written as
\begin{align}
	\label{Fried_Q}
	H^2_{\tilde{K}}(a)
	=
	\HdS^2\left[
		1
		+ \frac{Q_{\tilde{K}}^2}{\left(a\HdS \right)^{4-2n}}
	\right]
	\,,
\end{align}
where we have introduced the dimensionless parameter $Q_{\tilde{K}}$ as
\begin{align}
	\label{QK_def}
	Q_{\tilde{K}} := \frac{\lambda_*\HdS}{\left(a_*\HdS \right)^{n}}\,\tilde{K}
	\,.
\end{align}
We can solve this Friedmann equation analytically for both values of $n$.

\begin{figure}[t]
\centering
\includegraphics[width=.495\textwidth]{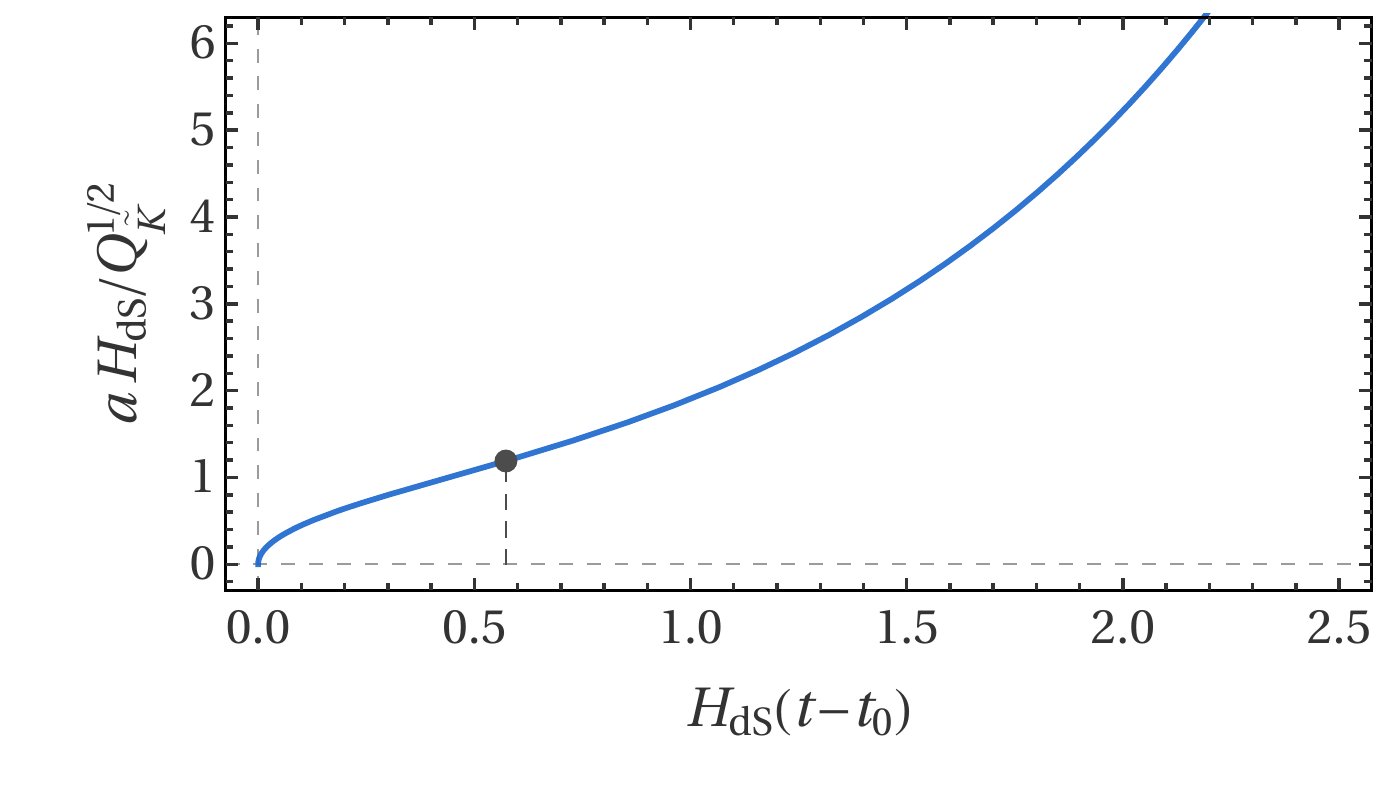}
\hfill
\includegraphics[width=.495\textwidth]{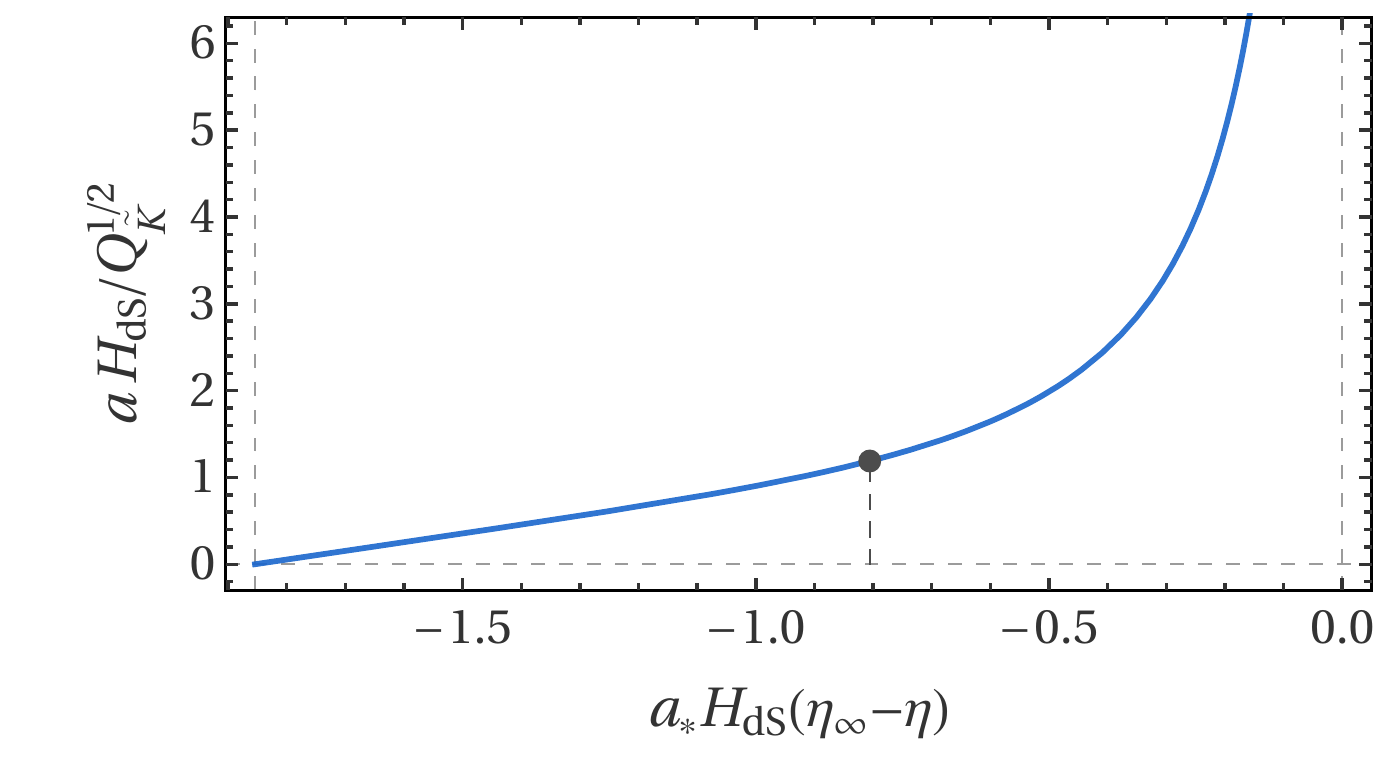}%
\vspace{-5pt}
\caption{\label{fig_evol_Rad}%
The evolution of the scale factor in terms of cosmic time in a universe with a radiation-like pre-inflationary phase (cf.~Eq.~\eqref{aradt}, left panel) and of the conformal time (cf.~Eq.~\eqref{aradeta}, right panel). The point in time where the decelerated expansion turns over to inflation is indicated by a dark dot.
}
\vspace{10pt}
\centering
\includegraphics[width=.495\textwidth]{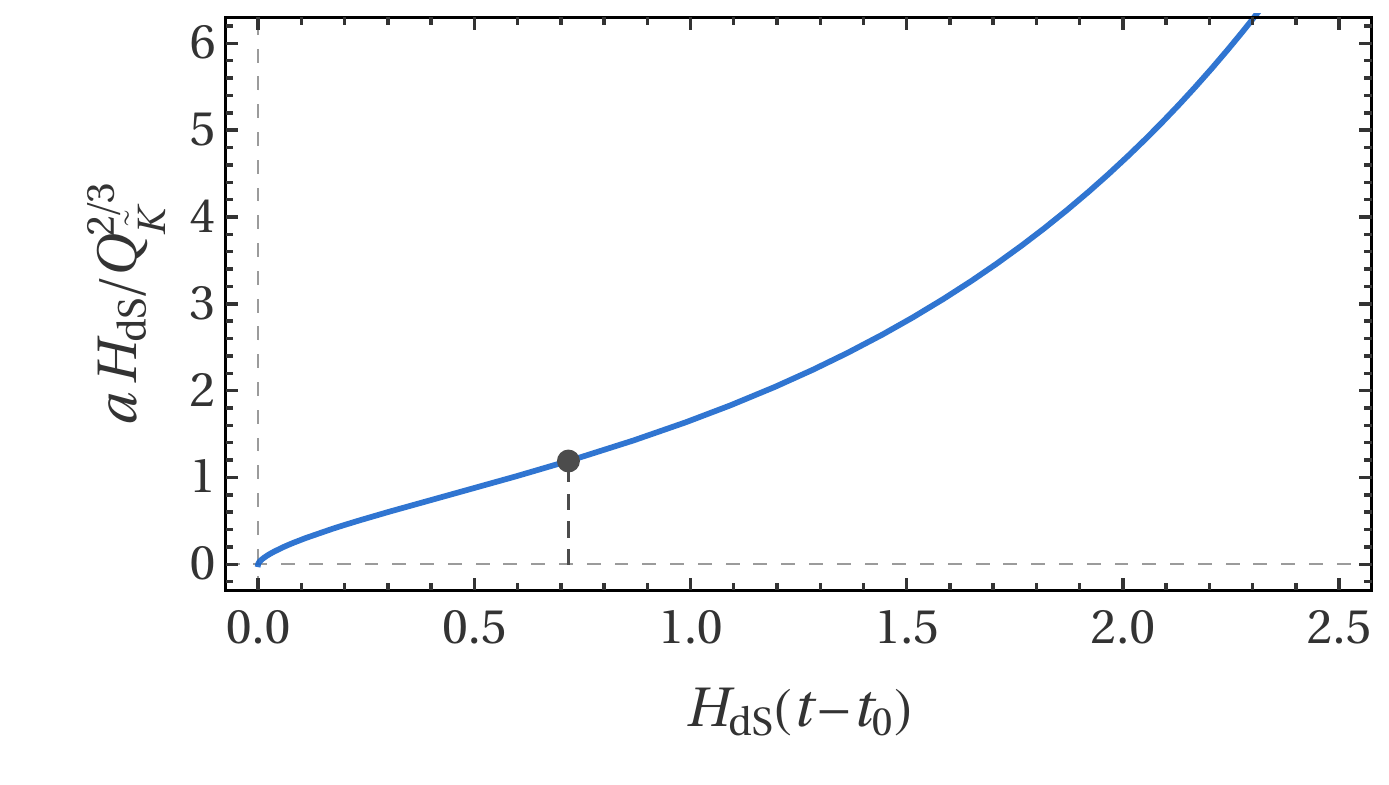}
\hfill
\includegraphics[width=.495\textwidth]{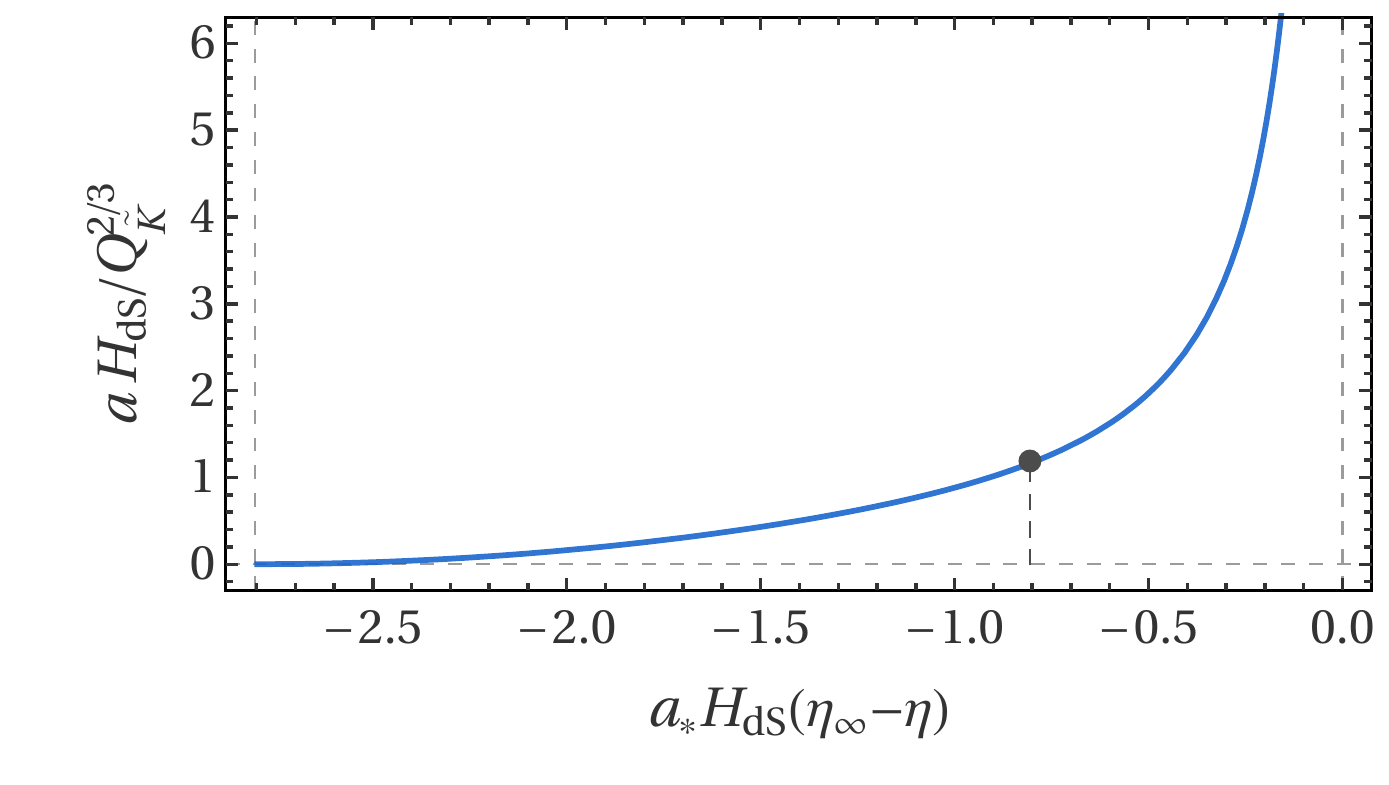}%
\vspace{-5pt}
\caption{\label{fig_evol_Dust}%
The evolution of the scale factor in terms of cosmic time in a universe with a dust-like pre-inflationary phase (cf.~Eq.~\eqref{adustt}, left panel) and of the conformal time (cf.~Eq.~\eqref{adusteta}, right panel). The point in time where the decelerated expansion turns over to inflation is indicated by a dark dot.} 
\end{figure}

For the radiation-like case $n=0$, we get the following evolution in terms of cosmic time
\begin{align} \label{aradt}
	a^2_{\tilde{K}}(t) = \frac{Q_{\tilde{K}}}{\HdS^2}\sinh\left[2\HdS(t-t_0)\right],
\end{align}
which is plotted in the left panel of Fig.~\ref{fig_evol_Rad}. In terms of the conformal time the solution in terms of the Jacobi elliptic function $\text{cn}$ reads \cite{Gradshteyn}
\begin{align} \label{aradeta}
	a^2_{\tilde{K}}(\eta) 
	=
	\frac{Q_{\tilde{K}}}{\HdS^2}\frac{1+\mathrm{cn}\!\left[2Q^{1/2}_{\tilde{K}}(\eta_\infty - \eta)\middle| k^2\right]}{1-\mathrm{cn}\!\left[2Q^{1/2}_{\tilde{K}}(\eta_\infty - \eta)\middle| k^2\right]}\,,\qquad k:=\frac{\sqrt{2}}{2}
\end{align}
and is shown in the right-hand panel of Fig.~\ref{fig_evol_Rad}. In the dust-like case $n=1/2$, we find the expressions
\begin{align} \label{adustt}
	a^{3/2}_{\tilde{K}}(t) = \frac{Q_{\tilde{K}}}{\HdS^{3/2}}\sinh\left[\frac{3}{2}\HdS(t-t_0)\right]
\end{align}
and
\begin{align} \label{adusteta}
	a^{3/2}_{\tilde{K}}(\eta) 
	=
	\frac{Q_{\tilde{K}}}{\HdS^{3/2}}\frac
	{1-\mathrm{cn}\!\left[{\sqrt[4]{3}}Q^{1/2}_{\tilde{K}}(\eta - \eta_0)\middle| k^2\right]}
	{(1+\sqrt{3})\,\mathrm{cn}\!\left[{\sqrt[4]{3}}Q^{1/2}_{\tilde{K}}(\eta - \eta_0)\middle| k^2\right] - (1-\sqrt{3})} 
	\,,\qquad k:=\frac{\sqrt{2+\sqrt{3}}}{2}\,.
\end{align}
These are plotted in the left and right panel of Fig.~\ref{fig_evol_Dust}, respectively.

\subsection{The scalar perturbations in the individual universes}

In order to derive the evolution equation for the quantized Mukhanov--Sasaki variables $\hat{v}_{k}$ from the Schr\"odinger equation \eqref{Schr_vk1}, we follow e.\,g.~\cite{bkk15,bkk16} and use a Gaussian ansatz for the wave function $\psi_{k,\tilde{K}}$ with normalization factor $N_{k,\tilde{K}}(\eta)$ and the function $\Omega_{k,\tilde{K}}(\eta)$:
\be
\label{Gauss}
\psi_{k,\tilde{K}}(\eta,v_k) =
N_{k,\tilde{K}}(\eta)\,\text{e}^{-\frac{1}{2}\,\Omega_{k,\tilde{K}}(\eta)\,\hat{v}_{k}^2}\,.
\ee
By using the relation
\be
\Omega_{k,\tilde{K}}(\eta) = -\,\I\,\frac{\hat{v}^\prime_{k}(\eta)}{\hat{v}_{k}(\eta)}\,,
\ee
we obtain the usual evolution equation of the quantized perturbation variables $\hat{v}_{k}$,
\be
\label{mode_evolution}
\hat{v}_{k}'' + \omega_{k,\tilde{K}}^2(\eta)\,\hat{v}_{k} = \hat{v}_{k}'' + \left(k^2 - \frac{z_{\tilde{K}}''}{z_{\tilde{K}}}\right) \hat{v}_{k} = 0\,,
\ee
where we have introduced $z_{\tilde{K}} := a(\varphi'/\mathscr{H}_{\tilde{K}})$. However, in order not to obfuscate the notation, we will skip the index $\tilde{K}$ for the quantity $z$.

We can now take a look at the different wave-number limits for Eq.~\eqref{mode_evolution}. For small wave numbers, $k^2\gg z''/z$, the solution to \eqref{mode_evolution} is given by the Bunch--Davies vacuum \cite{Mukhanov:1990me,Bassett:2005xm,Mukhanov:2005sc}
\begin{align}
	\label{ground_state}
	v_k^\text{BD} = \sqrt{\frac{1}{2k}}\,\E^{-\I k\eta}\,.
\end{align}
For lower wave numbers that fulfill $k^2<z''/z$, this solution is no longer valid and the amplitudes of the modes can be enhanced or suppressed instead of following an oscillatory behaviour like for the Bunch--Davies vacuum.

We can consider an equation of state (EoS) with a constant EoS parameter $w:=P/\rho$ and find that it is possible to solve the evolution equation \eqref{mode_evolution} analytically in terms of Hankel functions of the first and second kind \cite{abra,NIST2010} with order $\mu_w$ \cite{Bassett:2005xm,Contaldi:2003zv}:
\begin{align}
	\label{solutions_anal}
	v_{k,w} = 
	\frac{\sqrt{\pi|\eta|}}{2}\left[
		c_{1k} H_{\mu_w}^{(1)}\left(k|\eta|\right)
		+ c_{2k} H_{\mu_w}^{(2)}\left(k|\eta|\right)
	\right],
	\qquad
	\mu_w :=  
	\sqrt{2\dfrac{1-3w}{\left(1+3w\right)^2}
			+ \frac{1}{4}}
	\,.
\end{align}
Here%
\footnote{For $w=-1/3$ the potential $z''/z$ is constant, which leads to trivial oscillatory solutions for $v_k$, see e.g.~\cite{BouhmadiLopez:2012by}.},
we take $\eta>0$ for $w>-1/3$ and $\eta<0$ if $w<-1/3$. From the normalization condition
\begin{align}
	\label{normalization}
	v_k {v_k^*}'
	- v_k^* {v_k'}
	= \I
	\,,
\end{align}
we can infer that the linear coefficients $c_{1k}$ and $c_{2k}$ have to obey the constraint
\begin{align}
	\label{cik_normalization}
	|c_{1k}|^2 - |c_{2k}|^2 = \pm1\,,
\end{align}
where the upper positive sign in \eqref{cik_normalization} corresponds to inflationary phases with $w<-1/3$ while the lower negative sign corresponds to the cases of decelerated expansion with $w>-1/3$.

For more complicated equations of state, one cannot in general find analytic solutions. However, we see that, as we will show in the next section, analysing the evolution of $z''/z$ can help us to find out which modes are most affected during a given epoch of the universe's expansion.


In order to describe the anisotropies of the CMB, we need to introduce the primordial power spectrum of the comoving curvature perturbations $\mathcal{R}$, which are related to the Mukhanov--Sasaki variables by $v=z\mathcal{R}$ \cite{Lyth:1984gv,Mukhanov:1990me,Bassett:2005xm,Mukhanov:2005sc}. Thus the power spectrum can be written as \cite{Bassett:2005xm}
\begin{align}
	\label{power spectrum}
	P_{\mathcal{R}}(k) 
	= \frac{k^3}{2\pi^2}|\mathcal{R}_k|^2
	= \frac{k^3}{2\pi^2}\frac{|v_k|^2}{z^2}
	\,.
\end{align}
When describing CMB anisotropies, this power spectrum is usually fitted to a power law with amplitude $A_s$, spectral index $n_s$ and pivot scale $k_*$ in the following way
\begin{align}
	\label{power_spectrum_fit}
	P_{\mathcal{R}}(k) = A_s \left(\frac{k}{k_*}\right)^{n_s-1}\,.
\end{align}
The Planck mission uses a pivot scale of $k_* = 0.05$~$\mathrm{Mpc}^{-1}$ and in the 2018 Planck data, the preferred values for these parameters are 
$A_s =2.10732\times10^{-9}$ and $n_s=0.96824$, where also lensing effects and external BAO data have been used \cite{Akrami:2018vks}.

Going back to our example of a simple equation of state with parameter $w$, we can fix the linear coefficients $c_{ik}$ in \eqref{solutions_anal} up to an arbitrary non-physical phase by using both the normalization condition \eqref{cik_normalization} and by demanding that the Bunch--Davies vacuum \eqref{ground_state} is recovered in the low wave-length regime, i.e.~for $|k\eta|\gg1$.

Taking into account the asymptotic behaviour of the Hankel functions for large $k\eta$, we need to choose $c_{1k}=0$ and $|c_{2k}|=1$ for $w>-1/3$, whereas for $w<-1/3$, $|c_{1k}|=1$ and $c_{2k}=0$ need to be taken, such that we obtain
\begin{align}
	\label{Hankel_power}
	P_{\mathcal{R}}(k) 
	= \frac{k^3}{16\pi\epsilon} \frac{|\eta|}{a^2}\left|H_\mu^{(1)}(k|\eta|)\right|^2.
\end{align}
This expression arises from the fact that the two Hankel functions with a real-valued variable are complex conjugate to each other \cite{abra,NIST2010}.

In the long wave-number regime, for which the modes remain in the ground state, we can approximate the power spectrum in \eqref{Hankel_power}, independently of which value $w$ takes, as follows \cite{Bassett:2005xm}
\begin{align}
	P_{\mathcal{R}}(k|\eta|\gg1) 
	\simeq
	\frac{1}{\pi\epsilon}\frac{H^2}{\MP^2} \left(\frac{k}{a H}\right)^2.
\end{align}
However, for low wave numbers ($k|\eta|\ll1$), the primordial power spectrum explicitly depends on the specific expansion of the universe.
For an arbitrary parameter $w<1$ and specifically for near de Sitter inflation with $w=-1+\alpha/3\gtrsim-1$, we obtain
\begin{align}
	\label{spectrum_walpha}
	P_{\mathcal{R}}(k|\eta|\ll1) 
	\simeq
	\frac{2}{\pi} \frac{(2-\alpha)^{\frac{2\alpha}{2-\alpha}}}{\alpha}
	\left[
		{\left(1-\alpha/2\right)}
		\frac{\Gamma\left(\frac{1}{2}
		\frac{6-\alpha}{2-\alpha}\right)}{\Gamma(3/2)}
	\right]^2
	\frac{H^2}{\MP^2} 
	\left(\frac{k}{a H}\right)^{-\frac{2\alpha}{2-\alpha}}.
\end{align}


\section{Numerical results}
\label{Numerical}

\subsection{Description of the toy model}
\label{Sec_AsympSols}

In the model we described before, we assumed that the potential of the scalar field is quasi-constant (cf.~\eqref{Vmp}), which led to an effective de Sitter expansion of the universe after the pre-inflationary phase. However, in order to be able to describe the features of the measured CMB anisotropy spectrum on smaller scales, in particular the measured tilt of the power spectrum, we will now multiply the constant $\HdS^2$ in Eq.~\eqref{Fried_bg} by a term proportional to $a^{-\alpha}$ with a small parameter $\alpha\gtrsim0$, which corresponds to power-law inflation. That way, the desired tilt in the primordial power spectrum is introduced and we can obtain a spectral index $n_s$ that differs from one and is thus in agreement with observation. Hence, we end up with the following effective Friedmann equation
\begin{align}
	\label{bkgd_power_law}
	H^2_{\tilde{K}}(a)
	=
	\HdS^2\left[
		\left(\frac{a_*}{a} \right)^{\alpha}
		+ \frac{\lambda(a)^2\tilde{K}^2}{\HdS^2 a^4}
	\right]
	\,,
\end{align}
where $a_*$ corresponds to an arbitrary scale factor, which will be fixed later on.

As discussed in Sec.~\ref{section2}, for the moment, the interaction coupling $\lambda(a)$ is introduced in a purely phenomenological way \cite{Lievens2006,Lievens2007,Alonso2012} and as such we have the freedom to consider different choices for its functional form. Of particular interest is the case of a power-law dependence $\lambda(a)=\lambda_*(a/a_*)^n$ which can lead to different pre-inflationary epochs in the semi-classical regime or even an effective cosmological constant \cite{RP2016}.
The constant $\lambda_*$ defined in this way has dimensions of length.
In this work we will pay special attention to the cases $n=0$ and $n=1/2$ in which the second term inside the brackets on the right-hand side of Eq.~\eqref{bkgd_power_law} leads, respectively, to a radiation-like and dust-like initial epoch. Using the definition \eqref{QK_def} of the (dimensionless) rescaled $\tilde{K}$-number $Q_{\tilde{K}}$, we rewrite Eq.~\eqref{bkgd_power_law} as
\begin{align}
	\label{bkgd_power_law_param}
	H^2_{\tilde{K}}(a)
	=
	\HdS^2\left[
		\left(\frac{a_*}{a} \right)^{\alpha}
		+ \frac{Q_{\tilde{K}}^2}{\left(a\HdS\right)^{4-2n}}
	\right].
\end{align}
If we require that the $Q_{\tilde{K}}$-dependent term be negligible during inflation, $a\gtrsim a_*$, then we arrive at the constraint
\begin{align}
	\label{first_limit}
	Q_{\tilde{K}}\ll (a_*\HdS )^{2-n}
	\,.
\end{align}
As long as $n<1$, the effective Friedmann equation \eqref{bkgd_power_law_param} describes the evolution of a universe that first undergoes an initial phase whose EoS parameter is given by $w=(1-2n)/3$, before power-law inflation with an EoS parameter of $w=-1 + \alpha/3$ takes over. The scale factor at the point of transition from the initial epoch of decelerated expasion to the accelerated inflationary expansion takes the form
\begin{align}
	\label{transition}
	a_\mathrm{trans} 
	= 
	a_*
	\left(\frac{2-2n}{2-\alpha}\frac{Q_{\tilde{K}}^2}{\left(a_*\HdS \right)^{4-2n}}\right)^{\frac{1}{4-2n-\alpha}}
	\,,
\end{align}
and the squared comoving wave number $(aH)^2$ reaches its minimum value at:
\begin{align}
	\label{kmin_def}
	k^2_\mathrm{min}
	:= 
	\left(aH\right)^2_{a=a_\mathrm{trans}}
	=
	 \frac{4 - 2n - \alpha}{2-\alpha}
	\left(\frac{2-\alpha}{2-2n}\right)^{\frac{2-2n}{4-2n-\alpha}}
	\left(\frac{Q_{\tilde{K}}^2}{\left(a_*\HdS \right)^{4-2n}}\right)^{\frac{2-\alpha}{4-2n-\alpha}}
	\left(a_*\HdS \right)^2
	\,.
\end{align}
\begin{figure}[t]
\centering
\includegraphics[width=.495\textwidth]{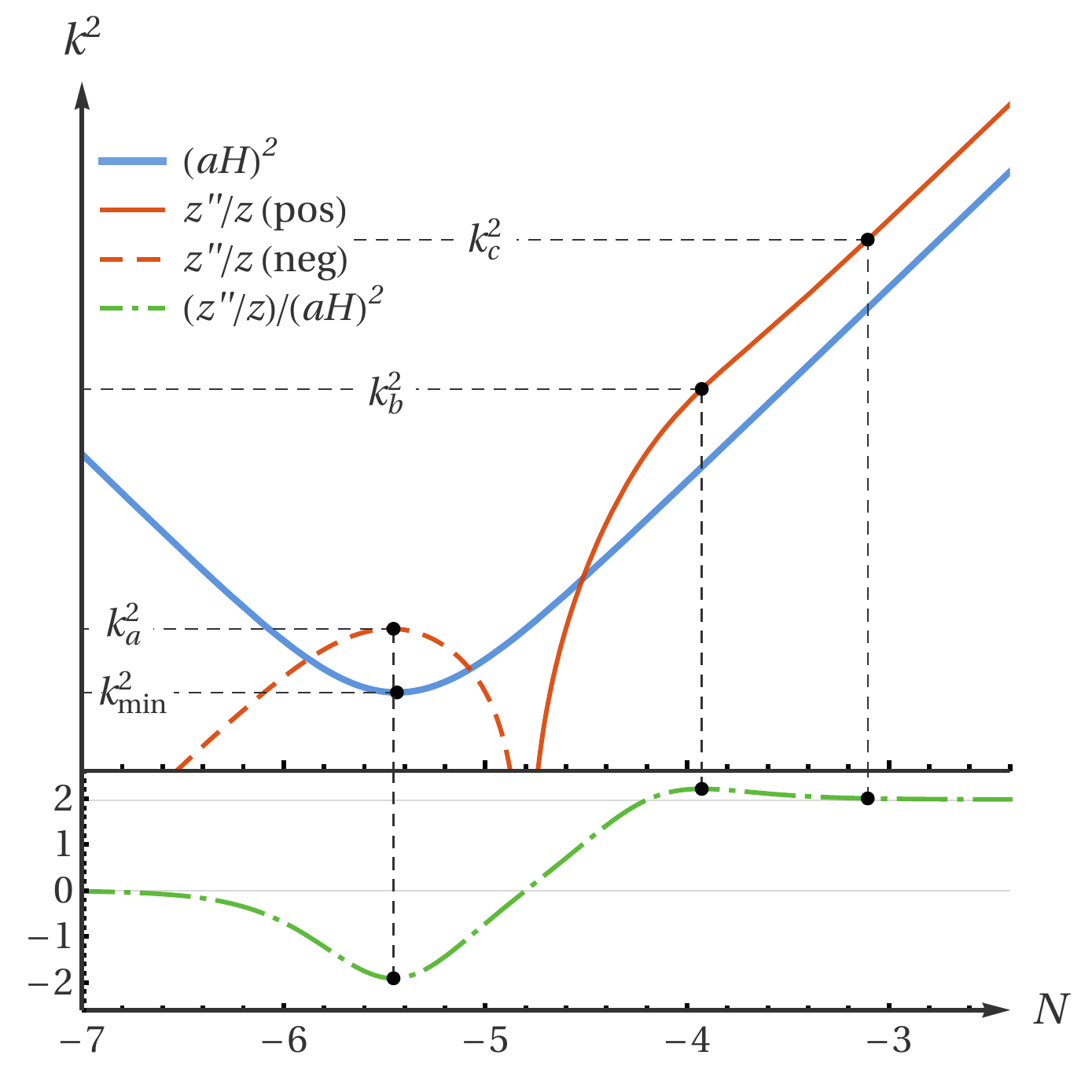}
\hfill
\includegraphics[width=.495\textwidth]{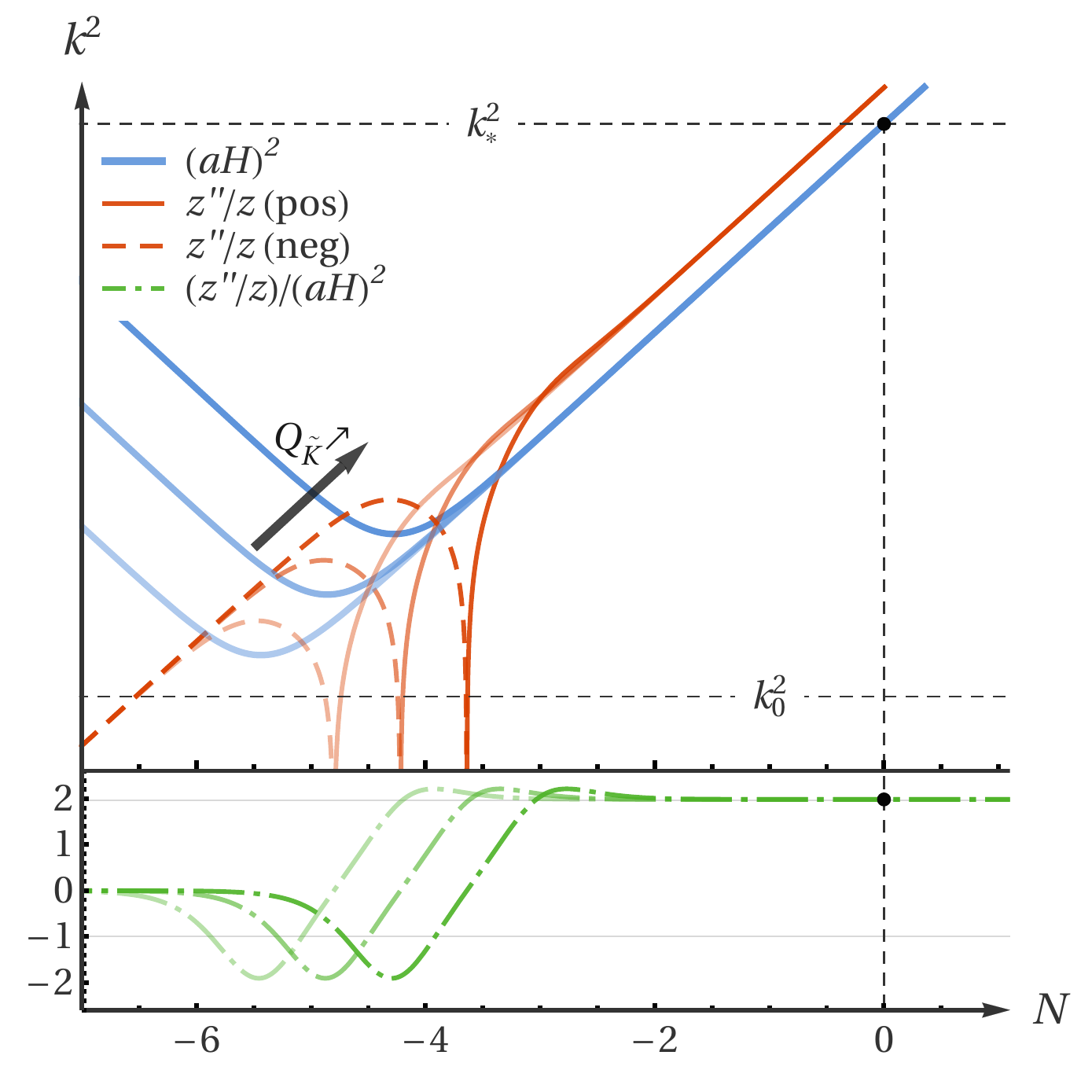}%
\vspace{-5pt}
\caption{\label{fig_potentials_Rad}%
(Left Panel) Characteristic shape of the squared Hubble horizon, $(aH)^2$, the scalar potential, $z''/z$, and the ratio $(z''/z)/(aH)^2$ around the transition from the pre-inflationary epoch to inflation for the model \eqref{bkgd_power_law} with $n=0$. The characteristic wave numbers $\ka$ and $\kb$ are defined based on the position of the local maximum and minimum of the ratio $(z''/z)/(aH)^2$. For $k>k_c=10\kmin$, no imprints of the model on the primordial power spectrum are expected.
(Right Panel) For higher values of the free parameter $Q_{\tilde{K}}$, the characteristic wave numbers $\kmin$, $\ka$, $\kb$ and $\kc$ increase and the imprints on the primordial power spectrum are shifted to scales closer to the pivot scale $k_*$. On the other hand, if $Q_{\tilde{K}}$ is so low that all the characteristic wave numbers are below $k_0$, no observable imprints are expected on the primordial power spectrum.}
\end{figure}
\begin{figure}[t]
\centering
\includegraphics[width=.495\textwidth]{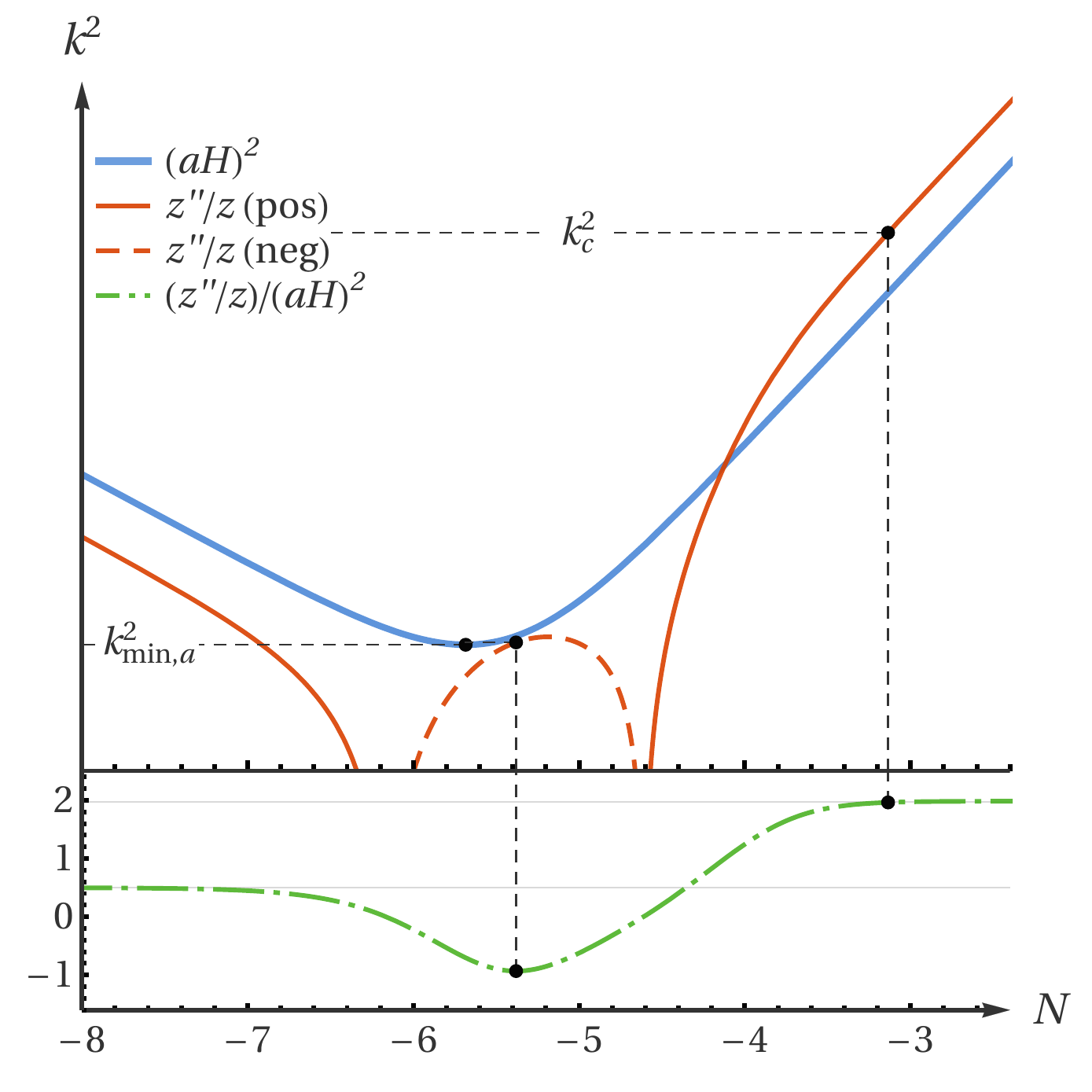}
\hfill
\includegraphics[width=.495\textwidth]{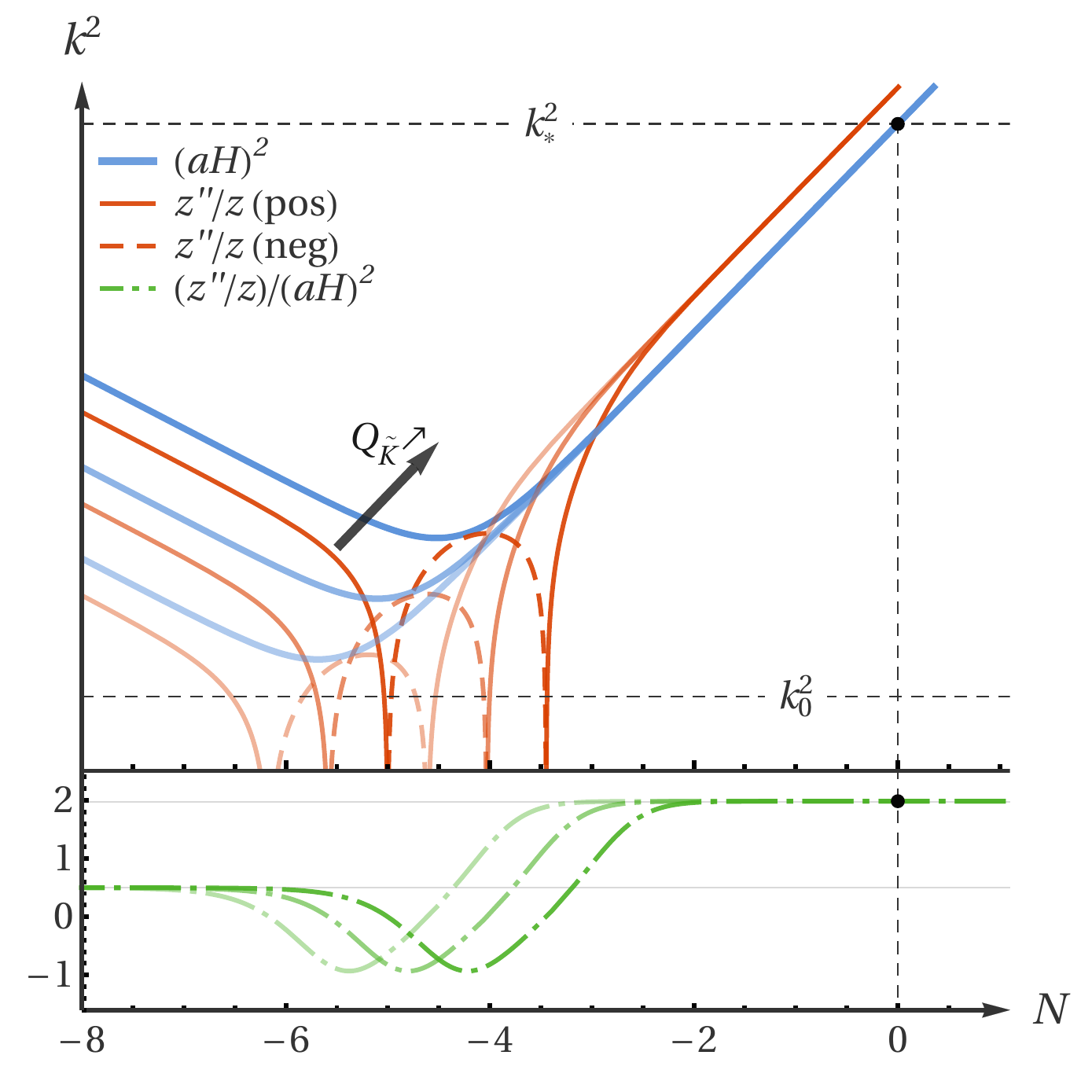}%
\vspace{-5pt}
\caption{\label{fig_potentials_Dust}%
(Left Panel) Characteristic shape of the squared Hubble horizon, $(aH)^2$, the scalar potential, $z''/z$, and the ratio $(z''/z)/(aH)^2$ around the transition from the pre-inflationary epoch to inflation for the model \eqref{bkgd_power_law} with $n=1/2$. The characteristic wave number is defined based on the position of the local  minimum of the ratio $(z''/z)/(aH)^2$. Since $\ka\approx\kmin$, in this case $\ka$ does not correspond to a new relevant scale to analyse the primordial power spectrum. The characteristic wave number $k_c=10\kmin$ sets the scale up to which imprints of the model on the matter power spectrum are expected.
(Right Panel) For higher values of the free parameter $Q_{\tilde{K}}$, the characteristic wave numbers $\kmin$, $\ka$ and $\kc$ increase, and the imprints on the primordial power spectrum should appear closer to the pivot scale $k_*$.}
\end{figure}

The evolution of the potential $z''/z$ (red), the wave number of the Hubble horizon (blue), as well as their ratio (green), is presented in Fig.~\ref{fig_potentials_Rad} for $n=0$ and in Fig.~\ref{fig_potentials_Dust} for $n=1/2$. 
In the asymptotic regions away from the transition, where $w$ is constant, the mode functions $v_k$ can be written as a linear combination of the solutions \eqref{solutions_anal}, with $\lambda=(1+n)/(2-2n)$ during the initial decelerating epoch and $\lambda=(6-\alpha)/(4-2\alpha)$ during  inflation.
This approximation is no longer valid during the transition and a numerical integration of the Mukhanov--Sasaki equation \eqref{mode_evolution} for each mode $k$ is required. 
Nevertheless, we can estimate the characteristic wave numbers which will contain imprints of the transition on the primordial power spectrum by looking at the shape of the potential $z''/z$ and the ratio $(z''/z)/(aH)^2$. 
More specifically, for the case of $n=0$ we define three characteristic wave numbers (cf.~left panel of Fig.~\ref{fig_potentials_Rad}):
\begin{itemize}
\item $k_\mathrm{a}$ and $k_\mathrm{b}$ correspond to the mode $k$ that crosses the potential $z''/z$ at the moment of the local minimum or maximum of $(z''/z)/(aH)^2$, respectively,
\item $\kc:=10\kmin$ corresponds to a mode that crosses the potential when $(z''/z)/(aH)^2$ has already reached (approximately) its asymptotic value of $2-\alpha/2$.
\end{itemize}
By numerical investigation we find that:
\begin{align}
	k_\mathrm{a} \approx&~ 1.382 \kmin\,,
	&
	k_\mathrm{b} \approx&~ 4.680 \kmin\,,
\end{align}
independently of the value of $Q_{\tilde{K}}$.
In the case of $n=1/2$, where no maximum appears in $(z''/z)/(aH)^2$, we define two characteristic wave numbers: $k_\mathrm{a}$ corresponds to the mode $k$ that crosses the potential $z''/z$ at the moment of the local minimum of $(z''/z)/(aH)^2$; and $\kc:=10\kmin$ corresponds to a mode that crosses the potential when $(z''/z)/(aH)^2$ has already reached (approximately) its asymptotic value of $2-\alpha/2$  (cf. left panel of Fig.~\ref{fig_potentials_Dust}). By numerical investigation, we find that $k_\mathrm{a} \approx 1.026 \kmin$, independently of the value of $Q_{\tilde{K}}$ and therefore it does not introduce a new relevant scale in the analysis of the shape of the primordial power spectrum of the model.

For $k>\kc$ we do not expect to see imprints on $P_\mathcal{R}$ of the pre-inflationary epoch and the transition to inflation.  Therefore, in order for the prediction of the model for the primordial power spectrum to respect the observational constraints%
\footnote{In the Planck 2018 final data release, several methods for a non-parametric reconstruction of the primordial power spectrum found no evidence for a deviation from a power law behaviour of ${P}_\mathcal{R}$ for scales between $0.005$ Mpc$^{-1}$ and $0.2$ Mpc$^{-1}$ \cite{Akrami:2018odb}.}, while at the same time not being red-shifted away to unobservable wave numbers, we can impose the constraint \mbox{$k_0\lesssim \kc \lesssim 10^{-1}k_*$}, where $k_0:=a_0 H_0$ is the wave number of the mode that is crossing the Hubble horizon at the present time. Using the definition \eqref{kmin_def}, this can be translated into the following constraints for the parameter $Q_{\tilde{K}}$:
\begin{align}
	\label{constraint1}
	& A_0 \left( \frac{k_0}{a_*\HdS }\right)^{\frac{4-\alpha}{2-\alpha}}
	\lesssim 
	\frac{Q_{\tilde{K}}}{\left(a_*\HdS \right)^2}
	\lesssim
	 A_0\left(\frac{k_*}{10a_*\HdS}\right)^{\frac{4-\alpha}{2-\alpha}}
	\,,
	&
	\textrm{for $n=0$}
	\,,
\end{align}
\begin{align}
	\label{constraint2}
	& A_{1/2} \left( \frac{k_0}{a_*\HdS }\right)^{\frac{3-\alpha}{2-\alpha}}
	\lesssim 
	\frac{Q_{\tilde{K}}}{\left(a_*\HdS \right)^{3/2}}
	\lesssim
	 A_{1/2}\left(\frac{k_*}{10a_*\HdS}\right)^{\frac{3-\alpha}{2-\alpha}}
	\,,
	&
	\textrm{for $n=1/2$}
	\,,
\end{align}
where
\begin{align}
	A_0 :=&~ \left[\frac{1}{10}\sqrt{{\frac{2-\alpha}{4-\alpha} }} \left(\frac{2}{2-\alpha}\right)^{\frac{1}{4-\alpha}}\right]^{\frac{4-\alpha}{2-\alpha}}
	\,,
	&
	A_{1/2} :=&~ \left[\frac{1}{10}\sqrt{{\frac{2-\alpha}{3-\alpha} }} \left(\frac{1}{2-\alpha}\right)^{{\frac{1}{2}}\frac{1}{3-\alpha}}\right]^{\frac{3-\alpha}{2-\alpha}}
	\,.
\end{align}
Under the assumption that the second term on the right-hand side of Eq.~\eqref{bkgd_power_law_param} is negligible for $a\gtrsim a_*$, we can use Eq.~\eqref{spectrum_walpha} at the moment of horizon crossing  to estimate the theoretical prediction of the primordial power spectrum for $k\gtrsim  a_*\HdS$:
\begin{align}
	\label{power_spectrum_approx}
	P_{\mathcal{R}}(k\gtrsim a_*\HdS)
	\simeq
	\frac{1}{2\pi\alpha}
	\left[
		\left(2-\alpha\right)^{\frac{2}{2-\alpha}}
		\frac{\Gamma\left(\frac{1}{2}
		\frac{6-\alpha}{2-\alpha}\right)}{\Gamma(3/2)}
	\right]^2 
	\frac{\HdS^2}{\MP^2}
	\left(\frac{k}{a_*\HdS}\right)^{-\frac{2\alpha}{2-\alpha}}
	\,.
\end{align}
By comparing this prediction with the fit to the  2018 data release of the Planck mission in combination with lensing effects and external BAO data \cite{Akrami:2018vks}, and using the freedom in choosing  the value of $a_*$ to fix $k_* = a_*\HdS$, we find that we can fix three of the four parameters of the model \cite{Morais2017}:
\begin{align}
	\label{observational_fit}
	\alpha \approx 0.031
	\,,
	\qquad
	\HdS
	\approx
	1.022 \times10^{-5}\, \MP
	\,,
	\qquad
	 a_*
	\approx
	2.167 \times10^{7}\, \MP^{-1}
	 \,.
\end{align}
With this normalization for the scale factor we have $a_0=H_0^{-1} \approx 8.454\times10^{60}\,\MP^{-1}$, where $a_0$ and $H_0$ are the current values of the scale factor and the Hubble rate, while $k_0:=a_0 H_0=1$ and $k_*\approx 221.4$.
Using the Planck values, the current day values of the Hubble rate and the scale factor are $H_0 \approx 1.184\times10^{-61}\,\MP$ and $ a_0=1\,\MP^{-1}$. Inserting \eqref{observational_fit} in the constraints \eqref{constraint1} and \eqref{constraint2} we find
\begin{align}
	\label{constraint1a}
	4.400 \times10^{-3}\, k_0^2~~
	\lesssim 
	& ~Q_{\tilde{K}}
	\lesssim
	2.266 \, k_0^2
	\,,
	&
	&\textrm{for $n=0$}
	\,,
	\\
	\label{constraint2a}
	1.838 \times10^{-2}\, k_0^{3/2}
	\lesssim 
	&~ Q_{\tilde{K}}
	\lesssim
	4.904 \times10^{-1}\, k_0^{3/2}
	\,,
	&
	&\textrm{for $n=1/2$}
	\,,
\end{align}
which are in conformity with the limit \eqref{first_limit}.
For these values  the number of e-folds of inflation before $a_*$ is
\begin{align}
	5.026 \lesssim &~ \log\left(\frac{a_*}{a_\mathrm{trans}}\right) \lesssim 8.173
	\,,
	&
	&\textrm{for $n=0$}
	\,,
	\\
	6.165 \lesssim &~ \log\left(\frac{a_*}{a_\mathrm{trans}}\right) \lesssim 8.378
	\,,
	&
	&\textrm{for $n=1/2$}
	\,.
\end{align}

\subsection{Numerical computations}

The strategy employed to solve the Mukhanov--Sasaki equation \eqref{mode_evolution}  for all the relevant modes follows the method used in Refs.~\cite{Henriques:1993km,Moorhouse:1994nc,Mendes:1994ai,BouhmadiLopez:2009hv,BouhmadiLopez:2011kw,BouhmadiLopez:2012by,BouhmadiLopez:2012qp} where the second-order linear differential equation for $v_k$ is replaced by a set of first-order linear {differential} equations for the variables $X_k= v_k$ and $Y_k=(\I/ k)X'$:
\begin{align}
	X_k' = -{\I k}Y_k
	\,,
	\qquad
	Y_k' = - \I k\left(1 - \frac{1}{k^2}\frac{z''}{z}\right)X_k
	\,.
\end{align}
The normalization constraint \eqref{normalization} can be written in terms of the real and imaginary parts of $X_k$ and $Y_k$ as
\begin{align}
	2k\left(
		X^\mathrm{(re)}_kY^\mathrm{(re)}_k 
		+ X^\mathrm{(im)}_kY^\mathrm{(im)}_k
	\right)
	=1
	\,.
\end{align}
This constraint is used in the numerical integrations to control the evolution of the numerical error.

The initial conditions for each numerical integration are set in a similar way to the one described in Ref.~\cite{Morais2017}:
\begin{itemize}
\item For the modes corresponding to $k<10\kc$, we impose the initial values of $X_k$ and $Y_k$ deep inside the initial decelerated period, at an initial moment $N=N_{\mathrm{ini},1}$ which is the same for all modes. The initial values of the integration variables are fixed using the solutions \eqref{solutions_anal} for $w=1/3$ (in the case $n=0$), or $w=0$ (in the case $n=1/2$), respectively, and by setting $c_{1k}=0$ and $c_{2k}=1$.
\item For the modes with $k>10\kc$, we assume that these modes are not influenced by the shape of the quantity $z''/z$ when going through the transition, such that we make the assumption that these modes are in the ground state when the inflationary phase starts. We set the initial values $X_k$ and $Y_k$ using the solutions \eqref{solutions_anal} with $w=-1+\alpha/3$, $c_{1k}=1$ and $c_{2k}=0$ at a certain number $ N_{\mathrm{ini},2}$ of e-folds before horizon crossing.
\end{itemize}
Notice that in comparison with Ref.~\cite{Morais2017}, here we set  a much lower bound for the transition between the two ranges of $k$ with different rules for initial conditions. This allows us to save computational time in the numerical integrations and is justified by the fact that the features in the potential are now much less pronounced when compared with the case studied in Ref.~\cite{Morais2017}%
\footnote{When obtaining the numerical results presented below, we have checked that changing the threshold wave number from $10^2\kc$ to $10\kc$ does not alter the primordial power spectrum obtained.}. The convergence of the numerical solutions is ensured by stopping the numerical integration at some fixed number of e-folds after horizon crossing during inflation. For modes with wave number $k<\kmin$ we stop the integration at the same moment as for $k=\kmin$ (see for example Fig.~3 of Ref.~\cite{Morais2017}).

Using the method aforementioned, we have performed eight numerical runs for each of the cases $n=0$ and $n=1/2$, and fixed the value of the free parameter $Q_{\tilde{K}}$ as
\begin{align}
	\label{tildeK_rad}
	Q_{\tilde{K},i}
	=&~
	k_0^2 10^{\frac{i}{2} - 3}
	\,,
	&
	\textrm{for }n=&~0
	\,,
	\\
	\label{tildeK_dust}
	Q_{\tilde{K},i}
	=&~
	k_0^{\frac{3}{2}} 10^{\frac{3}{4}\left(\frac{i}{2} - 3\right)}
	\,,
	&
	\textrm{for }n=&~1/2
	\,,
\end{align}
with $i=1,2,\dots,8$. We note that most of these values are below the upper limit defined in \eqref{constraint1a} and \eqref{constraint2a} as a way of guaranteeing that the near-scale-invariant shape of $P_\mathcal{R}$ around $k_*$ is recovered. As we will show below, in the numerical runs where $Q_{\tilde{K},i}$ is above those upper limits, imprints of the pre-inflationary epoch appear in  the primordial power spectrum for $k\gtrsim 10^{-1}k_*$, which in turn leads to modifications in $D_ \ell^{TT}$ for multipoles above $\ell=30$, thus validating our choice for the upper limits in this analysis. In addition to obtaining the theoretical prediction for the primordial power spectrum for each run we compute the normalized angular power spectra for the TT component \cite{Akrami:2018vks}, $D^{TT}_\ell:=\ell(\ell+1)C^{TT}_\ell/(2\pi)$, by feeding our numerical results  into the code CLASS \cite{CLASS}. The baseline 6-parameter $\Lambda$CDM model is assumed for the late-time cosmology and the values of its parameters are the ones found from the best fit to the 2018 data release of the Planck mission in combination with lensing effects and external BAO data \cite{Akrami:2018vks}.

\subsubsection{Multiverse with a radiation-like pre-inflation}

\begin{figure}[t]
\centering
\includegraphics[width=.495\textwidth]{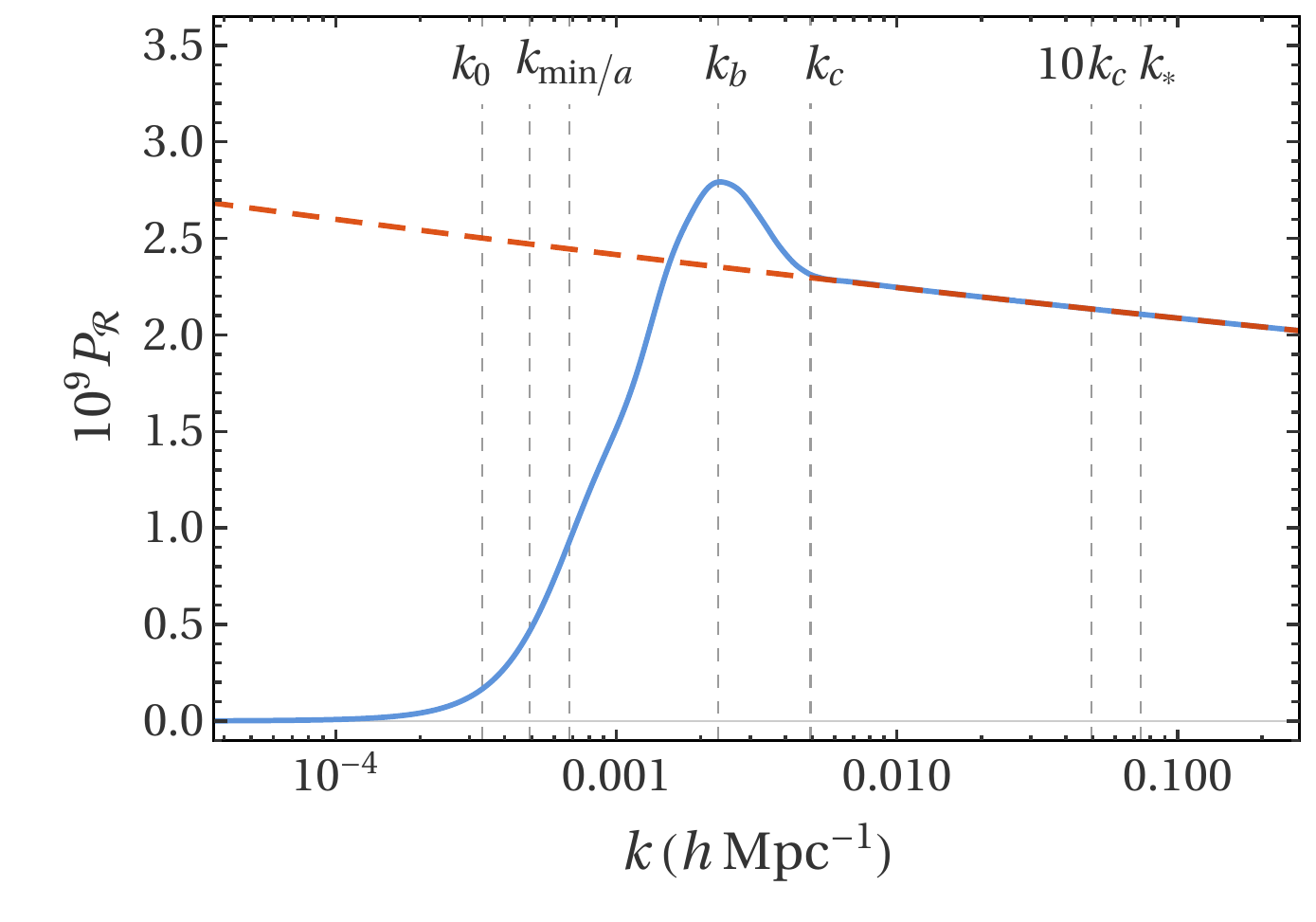}
\hfill
\includegraphics[width=.495\textwidth]{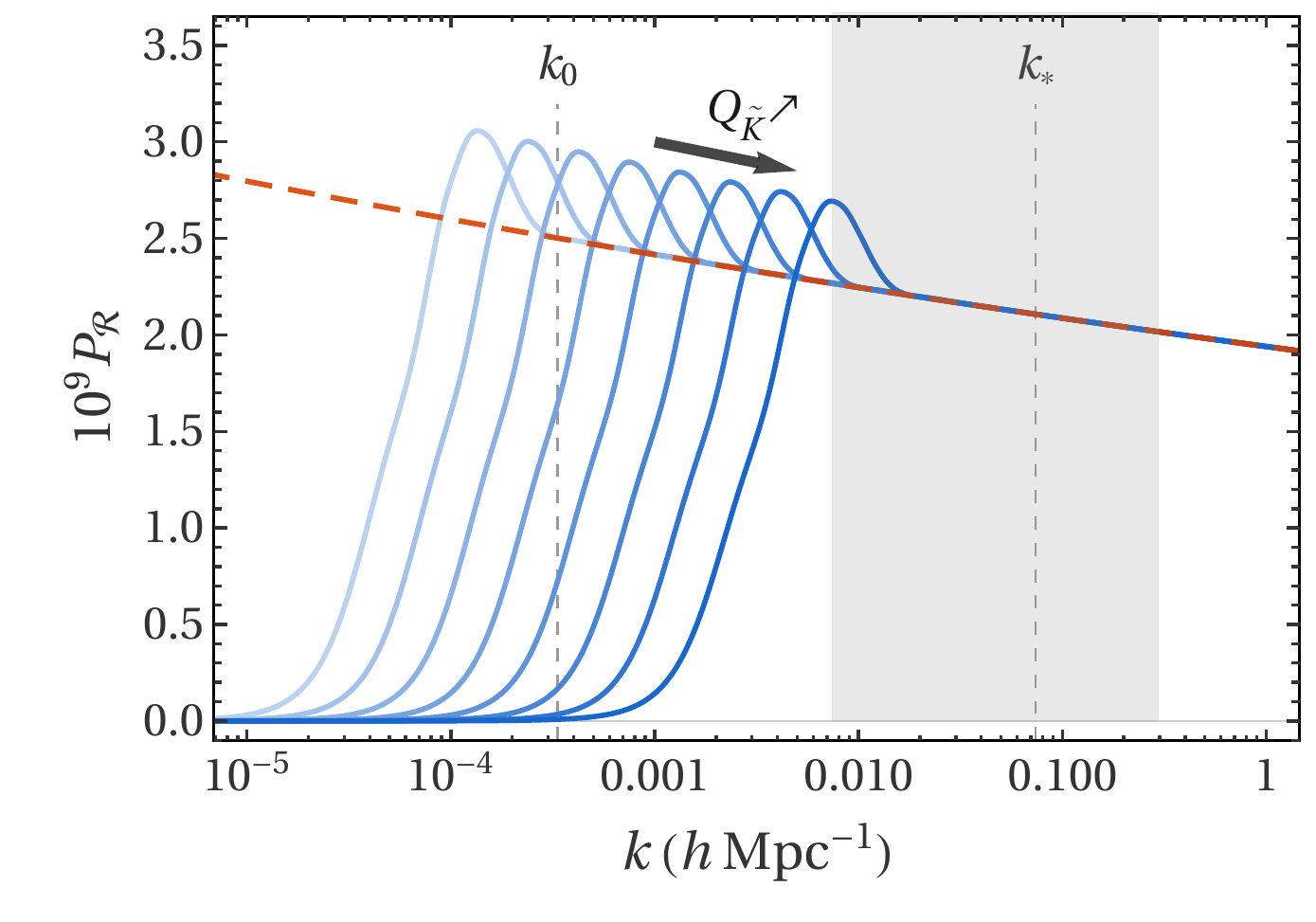}%
\vspace{-5pt}
\caption{\label{fig_spectra_Rad}%
(Left Panel) Characteristic shape primordial power spectrum for the model \eqref{bkgd_power_law} with $n=0$ compared with the prediction of the Planck best fit. The characteristic wave numbers defined based on the shape of the scalar potential $z''/z$ correspond to specific imprints on the power spectrum. No imprints of the model appear for $k>\kc$.
(Right Panel) As the value of the free parameter $Q_{\tilde{K}}$ increases, the imprints of the model on the primordial power spectrum are shifted to the right. For the values defined in \eqref{tildeK_rad}, the near-scale-invariant shape of the power spectrum is recovered for $k\gtrsim k_*$. The shaded region indicates the interval of wave numbers where no deviation from a power-law behaviour was found in Planck 2018 \cite{Akrami:2018odb}.}
\vspace{10pt}
\centering
\includegraphics[width=.495\textwidth]{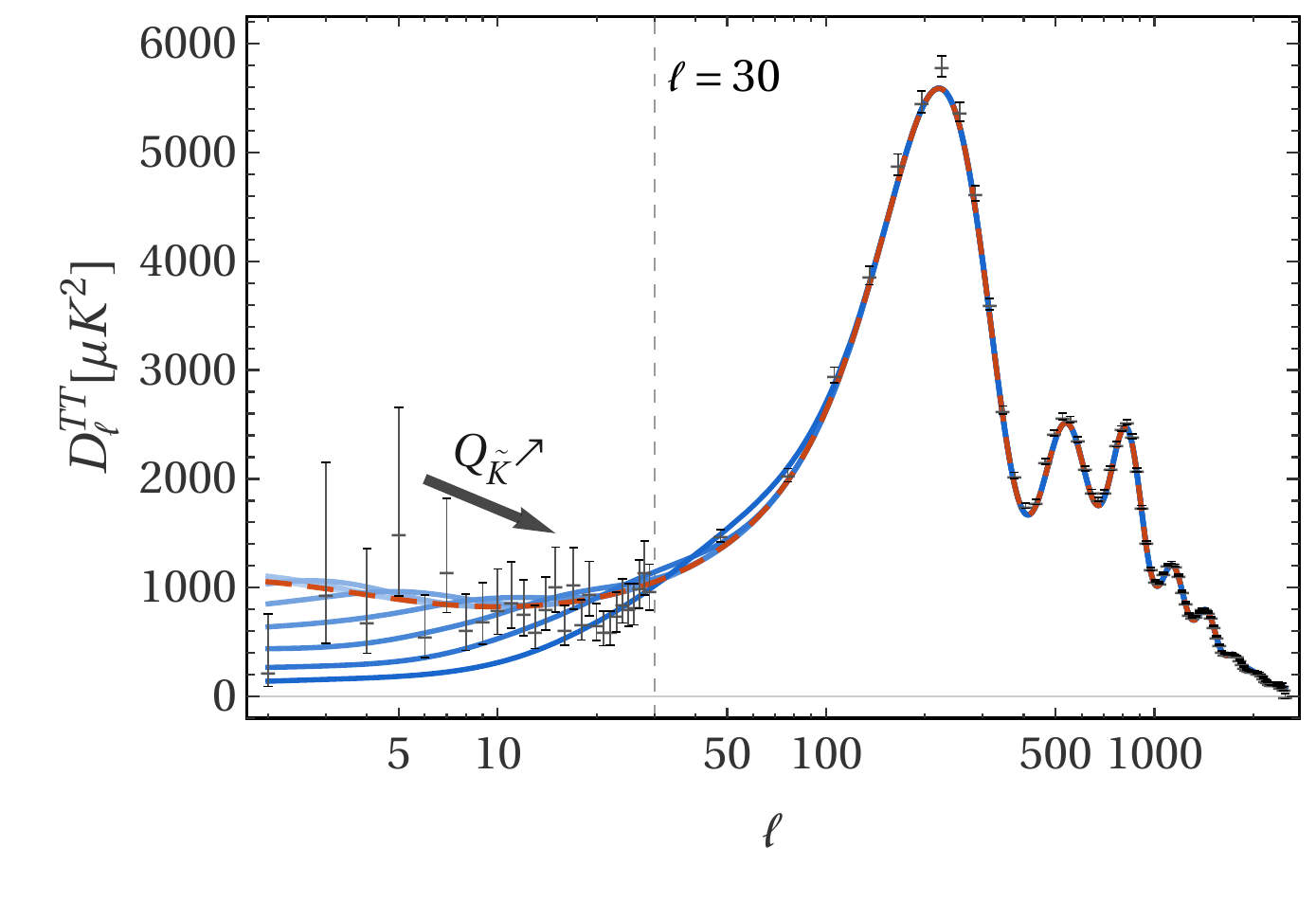}
\hfill
\includegraphics[width=.495\textwidth]{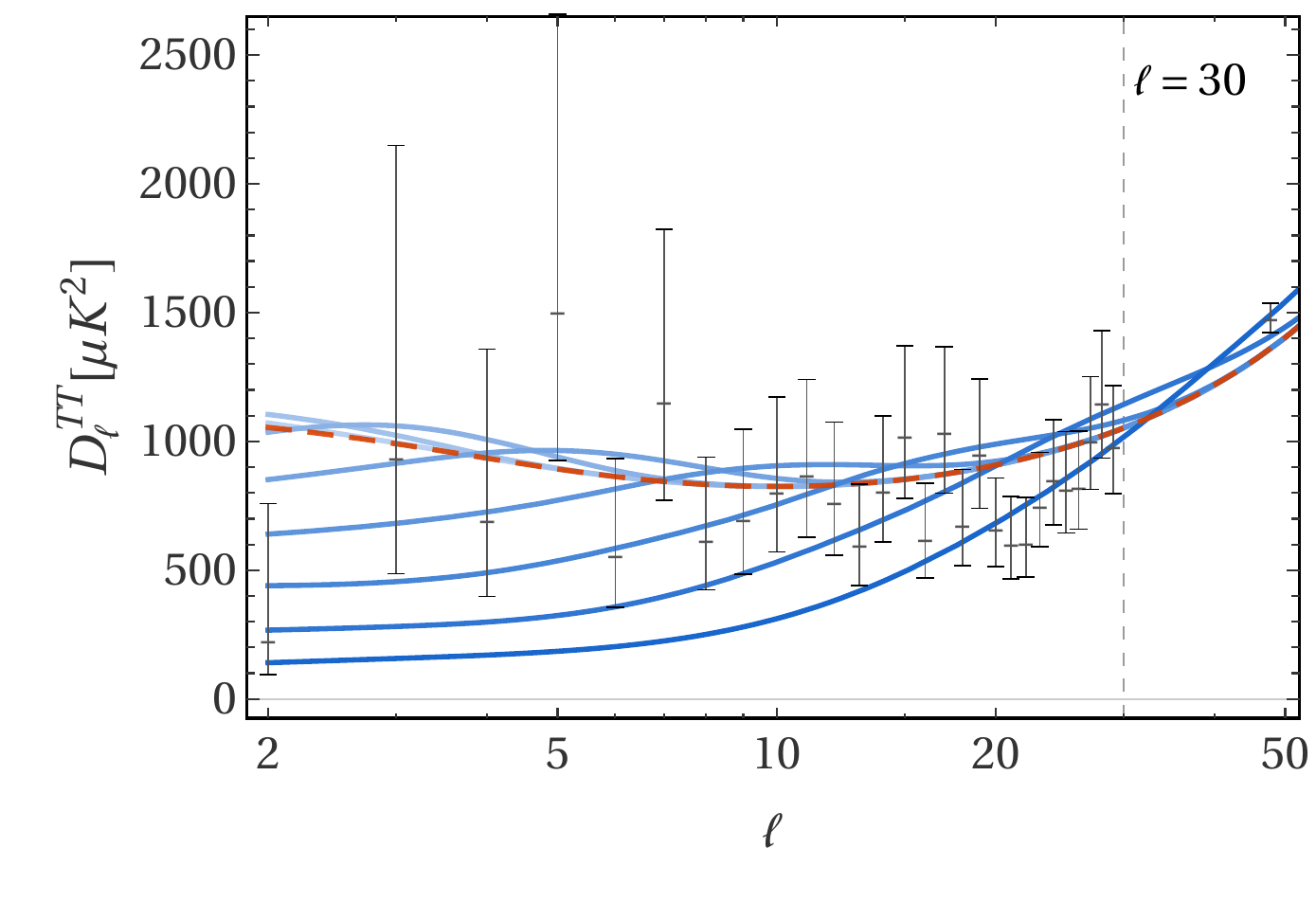}%
\vspace{-5pt}
\caption{\label{fig_Cl_Rad}%
(Left Panel) The angular power spectra $D_\ell^{TT}$ obtained from the primordial power spectra shown in Fig.~\ref{fig_spectra_Rad} compared with the prediction of the Planck best fit and the observational data points with respective error bars. A strong suppression is observed at the lowest multipoles, followed by a small peak. At high multipoles, all the curves converge to the prediction of the Planck best fit. 
(Right Panel) The angular power spectra $D_\ell^{TT}$ zoomed in the low multipole range $\ell\in(2,50)$.
As the value of the free parameter $Q_{\tilde{K}}$ increases, the peak observed in $D_\ell^{TT}$ is shifted to the right and suppression affects increasingly high multipoles. The two curves with rightmost peaks affect the shape of the angular power spectrum for $\ell>30$.}
\end{figure}

The results of the numerical integration for the case of $n=0$ are presented in Fig.~\ref{fig_spectra_Rad}, where on the left-hand side panel we plot the characteristic shape of the primordial power spectrum and on the right-hand side panel we show the effect of changing the value of the parameter $Q_{\tilde{K}}$. As can be seen on the left-hand side panel, the modes $\ka$, $\kb$ and $\kc$ defined in the previous section through the analysis of the shape of the potential $z''/z$ correspond to characteristic imprints in the primordial power spectrum. In particular, we observe a knee in the spectrum for  $k\approx\ka$ and for $k\approx\kb$ the spectrum presents a peak whose magnitude is well above the Planck best fit: $P_\mathcal{R}\approx 1.187 P_\mathcal{R}^\mathrm{Planck}$.
Notice, however,  that this peak is fairly small when compared to the results found in \cite{Morais2017} where a pre-inflationary era with $w=1$ was considered.
For $k\lesssim\ka$ we find that predicted primordial power spectrum is highly suppressed -- for $k\approx \kmin$ we find that $P_\mathcal{R}\approx 0.190 P_\mathcal{R}^\mathrm{Planck}$. On the other hand, for $k>\kc$ the usual near-scale-invariant shape of the primordial power spectrum is recovered, ensuring the agreement with observational data at low scales. We point out that for modes with $k\approx10\kc$ no special imprints are found, which validates the method described above to impose the initial conditions.

In order to demonstrate the effects of changing the parameter $Q_{\tilde{K}}$, we show on the right-hand side panel of Fig.~\ref{fig_spectra_Rad} the results of the numerical integration for the different values defined in Eq.~\eqref{tildeK_rad}. The lightest blue curve corresponds to $Q_{\tilde{K},1}$, while the darkest one corresponds to $Q_{\tilde{K},8}$.
As expected, as $Q_{\tilde{K}}$ increases, the imprints of the pre-inflationary era on the primordial power spectrum are shifted to higher wave numbers. Nevertheless, in most cases the peaks in Fig.~\ref{fig_spectra_Rad} only affect wave numbers below $10^{-1}k_*$; in the run with the highest value of the free parameter, $Q_{\tilde{K},8}=10k_0^2$, the peak appears at $k_b\approx10^{-1}k_*$. We note that $Q_{\tilde{K},1}$ is below the lower limit defined in \eqref{tildeK_rad}, consequently  the imprints on $P_\mathcal{R}$ in the observable range ($k>k_0$) are minimal.

In Fig.~\ref{fig_Cl_Rad} we present the normalized angular power spectra for the TT component of the CMB \cite{Akrami:2018vks}. On the left-hand side panel we plot the full spectra in the range $\ell\in(2,\,2500)$ and on the right-hand side panel we present a zoom of the spectra in the low multipole range $\ell\in(2,\,50)$. The theoretical predictions for the eight numerical runs, in blue, are contrasted against the available Planck 2018 data points and error bars%
\footnote{The data points and respective error bars were retrieve from the publicly available data in the Planck Legacy Archive \href{https://pla.esac.esa.int/pla/\#home}{https://pla.esac.esa.int/pla/}. Points in the range $2\leq\ell\leq29$ correspond to estimates from the Planck Commander algorithm, while for $\ell\geq 30$ the Planck Plik-lite binned estimates were used.}
and the prediction using the best-fit values of the cosmological parameters \cite{Akrami:2018vks} (red dashed curve).
As in Fig.~\ref{fig_spectra_Rad}, lighter blue curves in Fig.~\ref{fig_Cl_Rad} correspond to lower values of $Q_{\tilde{K}}$, cf.~\eqref{tildeK_rad}. Notice that for the cases $i=1,2$, for which $P_\mathcal{R}$ presents almost no special features for $k>k_0$, the angular power spectra in Fig.~\ref{fig_Cl_Rad} have minimal deviations with regard to the Planck fit. For the other cases we find that in the  low multipoles there is a suppression of $D_\ell^{TT}$, in accordance with the suppression observed in the primordial power spectrum. All the curves tend to follow the dashed curve corresponding to the Planck fit for higher multipoles. In an intermediate region we observe a small peak, whose position is $Q_{\tilde{K}}$-dependent. By requiring that this peak does not affect multipoles with $\ell\gtrsim30$, we can set the upper limit 
\begin{align}
	\label{upperlimit_radiation}
	Q_{\tilde{K}}\lesssim k_0^2\approx 2.042\times10^{-4}k_*^2
	\,, 
\end{align}
which is well inside the interval defined above in Eq.~\eqref{constraint1a}.

In order to determine which of the parameters $Q_{\tilde{K},i}$ fits the CMB data best, we calculate the reduced $\chi^2$-values taking into account only the CMB data points $(D_\ell^{TT})_\text{obs}$ and their measurement errors $\sigma_{\text{obs},\ell}$ from the Planck 2018 data up to $\ell = 30$ and comparing them to the $(D_\ell^{TT})_{\text{model},i}$ arising from our model for the eight values of $Q_{\tilde{K},i}$ that we chose: 
\be \label{chi2}
	\chi^2_{\text{red},i} 
	=
	\frac{1}{28}
	\sum_{\ell=2}^{29}\frac{1}{\sigma_{\text{obs},\ell}^2}
	\left[(D_\ell^{TT})_{\text{model},i} - (D_\ell^{TT})_\text{obs}\right]^2
	\,,
\ee
where the number in the denominator arises from the number of data points, 29, minus the number of free parameters, 1.
The $\chi^2$-values are listed in Table \ref{tabchi} on the left-hand side. As one can see, the value
\be \label{bestfit0}
	Q_{\tilde{K},5}
	\simeq 0.316\,k_0^2
	\simeq 6.458 \times 10^{-6} k_*^2
\ee
fits the CMB data best.

\setlength{\tabcolsep}{0.5em}
\begin{table}
\begin{center}
\begin{tabular}{  c  l  c  }
 \multicolumn{3}{c}{$n=0$ (radiation-like)} \\
  \hline	
  \hline		
  $i$ & $Q_{\tilde{K},i}/k_0^2$ & $\chi_{\text{red},i}^2$ \\
  \hline	
  1 & $~~10^{-2.5}$	&	1.186 \\	
  2 & $~~10^{-2}$	&	1.216  \\ 	
  3 & $~~10^{-1.5}$	&	1.169 \\
  4 & $~~10^{-1}$ 	&	1.074 \\
  5 & $~~10^{-0.5}$	& 	1.037 \\
  6 & $~~10^{0}$ 	& 	1.239 \\
  7 & $~~10^{0.5}$ 	& 	1.130 \\
  8 & $~~10^{1}$	&	1.297 \\
  \hline  
\end{tabular}
\qquad\qquad
\begin{tabular}{ c l l }
\multicolumn{3}{c}{$n=1/2$ (dust-like)} \\
  \hline	
  \hline
  $i$ & $Q_{\tilde{K},i}/k_0^{3/2}$ & $\chi_{\text{red},i}^2$ \\
  \hline		
  1 & $~~10^{-1.875}$	& 	1.158 \\	
  2 & $~~10^{-1.5}$ 	& 	1.098 \\ 	
  3 & $~~10^{-1.125}$ 	& 	0.983 \\
  4 & $~~10^{-0.75}$ 	& 	0.872 \\
  5 & $~~10^{-0.375}$ 	& 	0.823 \\
  6 & $~~10^{0}$ 		& 	0.845 \\
  7 & $~~10^{0.375}$ 	& 	1.421 \\
  8 & $~~10^{0.75}$ 	& 	3.310 \\
  \hline
\end{tabular}
\caption{The reduced $\chi^2$-values for the eight different choices of $Q_{\tilde{K},i}$ for the cases of a radiation-like (left-hand side) and a dust-like (right-hand side) pre-inflationary phase.}
\label{tabchi} 
\end{center}
\end{table}

\subsubsection{Multiverse with a dust-like pre-inflation}

\begin{figure}[t]
\centering
\includegraphics[width=.495\textwidth]{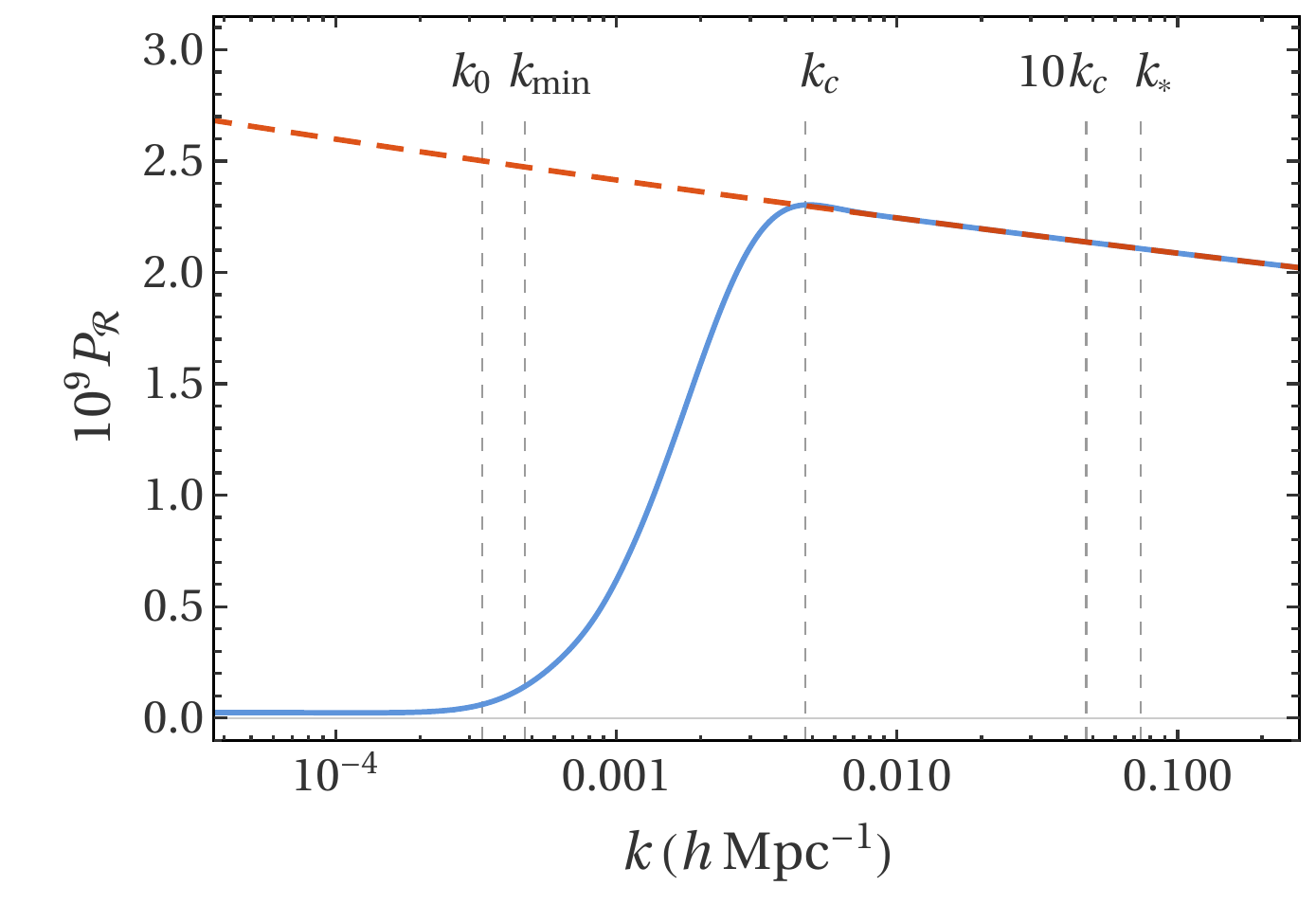}
\hfill
\includegraphics[width=.495\textwidth]{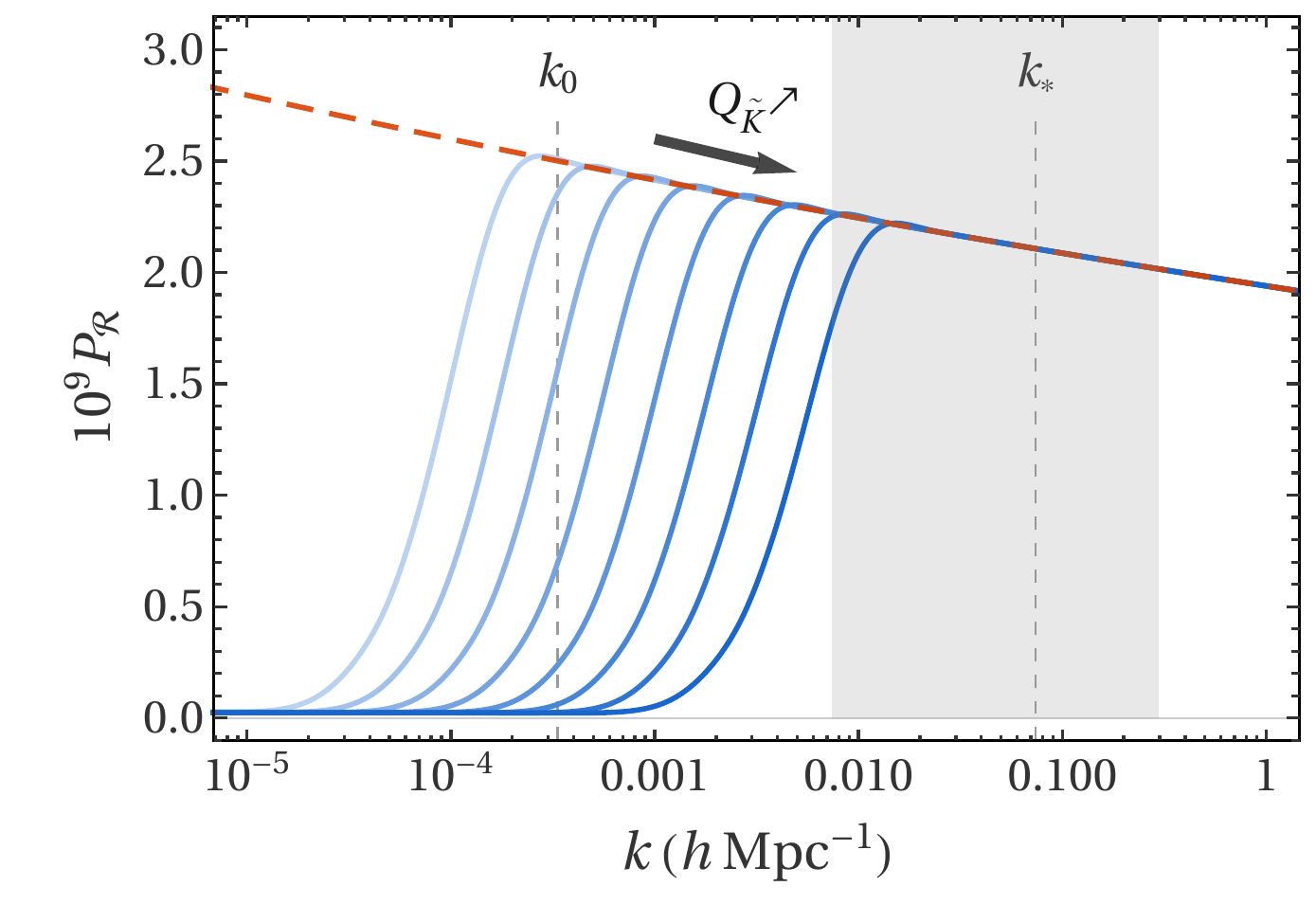}%
\vspace{-5pt}
\caption{\label{fig_spectra_Dust}%
(Left Panel) Characteristic shape primordial power spectrum for the model \eqref{bkgd_power_law} with $n=1/2$ compared with the prediction of the Planck best fit. The characteristic wave numbers defined based on the shape of the scalar potential $z''/z$ correspond to specific imprints on the power spectrum. No imprints of the model appear for $k>\kc$.
(Right Panel) As the value of the free parameter $Q_{\tilde{K}}$ increases, the imprints of the model on the primordial power spectrum are shifted to the right. For the values defined in \eqref{tildeK_rad}, the near-scale-invariant shape of the power spectrum is recovered for $k\gtrsim k_*$. The shaded region indicates the interval of wave numbers where no deviation from a power-law behaviour was found in Planck 2018 \cite{Akrami:2018odb}.}
\vspace{10pt}
\centering
\includegraphics[width=.495\textwidth]{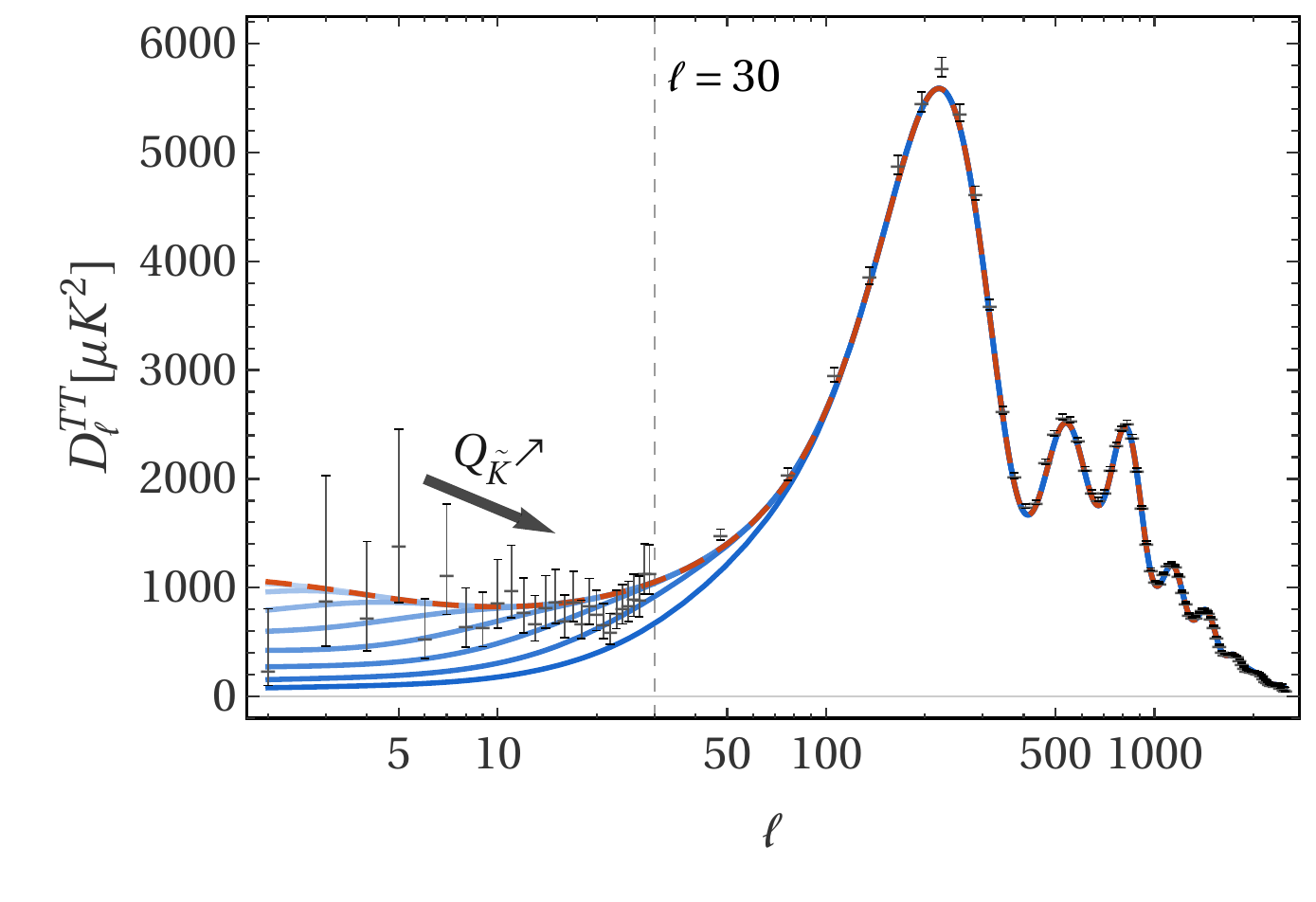}
\hfill
\includegraphics[width=.495\textwidth]{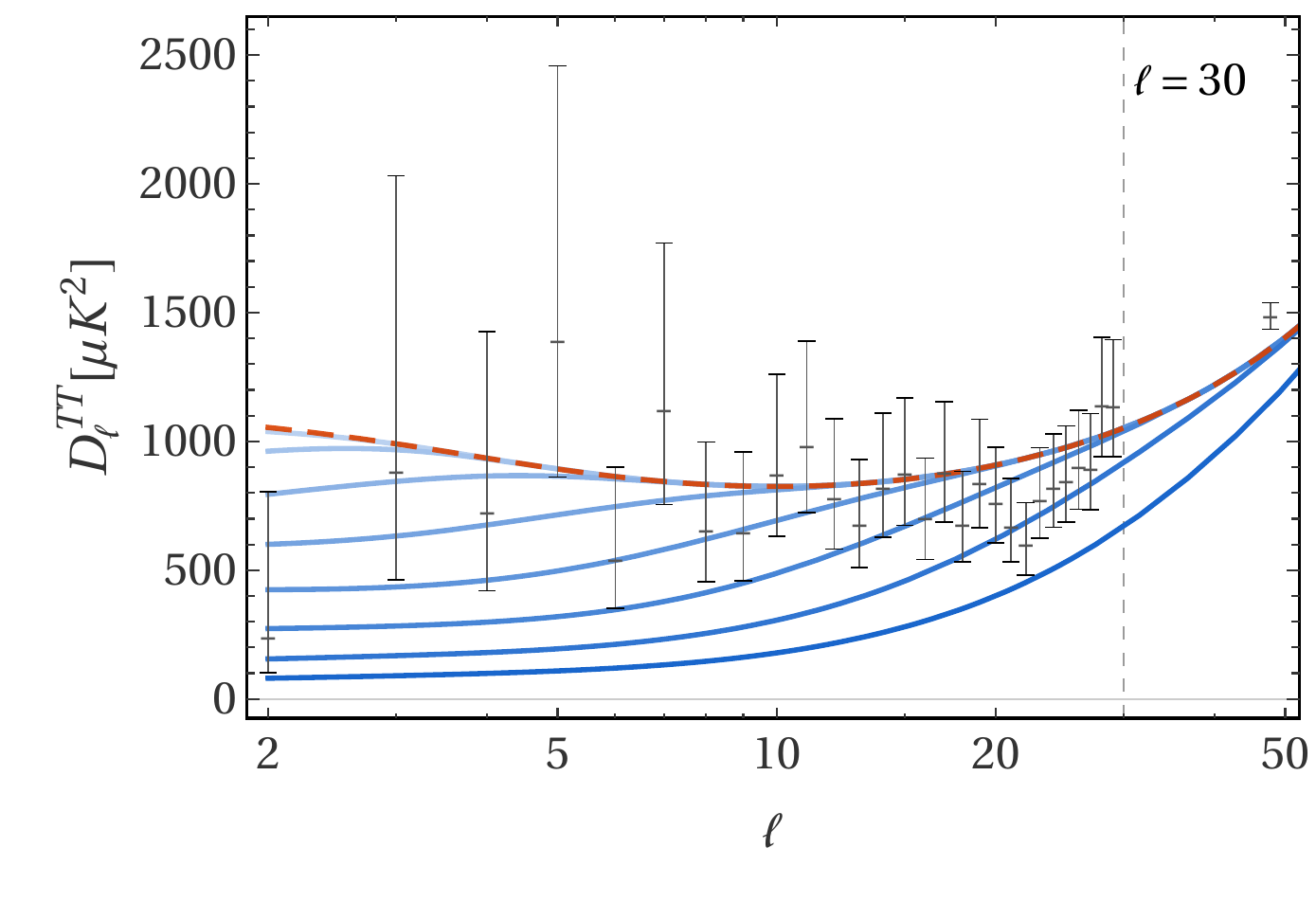}%
\vspace{-5pt}
\caption{\label{fig_Cl_Dust}%
(Left Panel) The angular power spectra $D_\ell^{TT}$ obtained from the primordial power spectra shown in Fig.~\ref{fig_spectra_Dust} compared with the prediction of the Planck best fit and the observational data points with respective error bars. A strong suppression is observed at the lowest multipoles, followed by a small peak. At high multipoles, all the curves converge to the prediction of the Planck best fit. 
(Right Panel) The angular power spectra $D_\ell^{TT}$ zoomed in the low multipole range $\ell\in(2,50)$.
As the value of the free parameter $Q_{\tilde{K}}$ increases, the peak observed in $D_\ell^{TT}$ is shifted to the right and suppression affects increasingly high multipoles. The two curves with rightmost peaks affect the shape of the angular power spectrum for $\ell>30$.} 
\end{figure}

In the case of $n=1/2$, we present the results of the numerical integration in Fig.~\ref{fig_spectra_Dust} where on the left-hand side panel we show the characteristic shape of the primordial power spectrum. We find that $P_\mathcal{R}$ follows the Planck best fit (dashed curve) for $k>\kc$, while observing a strong suppression at the larger scales -- we find that the predicted power spectrum is about $P_\mathcal{R}\approx 0.058 P_\mathcal{R}^\mathrm{Planck}$ at $k\approx\kmin$. At the characteristic scale $\kc$ the spectrum presents an almost indistinguishable peak. When comparing this result with the previous case of $n=0$, or with the results of Ref.~\cite{Morais2017}, we observe a tendency for the amplitude of the peaks  in the power spectrum to become smaller as the EoS parameter in the pre-inflationary epoch decreases. A similar effect is seen for example in Ref.~\cite{Cicoli2014} where instantaneous transitions are considered.
As in the previous case,  for modes with $k\approx10k_c$ we do not observe any deviation from the Planck fit for $P_\mathcal{R}$.

On the right-hand side panel of Fig.~\ref{fig_spectra_Dust}, we show the results of the numerical integration for the eight runs defined in \eqref{tildeK_dust}.
As $Q_{\tilde{K}}$ increases, the characteristic shape of the spectrum remains the same while the spectrum as a whole is shifted to higher wave numbers. Since all values $Q_{\tilde{K},i}$ are well below the upper limit in \eqref{constraint2a}, all the spectra in Fig.~\ref{fig_spectra_Dust}  follow the Planck best fit for  $k>10^{-1}k_*$.

Finally, in Fig.~\ref{fig_Cl_Dust} we present the angular power spectra  $D^{TT}_\ell$ obtained with the CLASS code. The theoretical predictions for $D^{TT}_\ell$ for the eight numerical runs, in blue, are plotted against the available Planck 2018 data points and error bars and  the predicted $D_\ell^{TT}$ obtained by using the best-fit values of the cosmological parameters \cite{Akrami:2018vks} (red dashed curve). The left-hand side panels shows the spectra for the entire range $\ell\in(2,2500)$, while the right-hand side panel corresponds to the zoom-in of the same figure on the low multipole range.
As in the previous figures, the blue curves in Fig.~\ref{fig_Cl_Dust} are colour-coded with regard to the value of the parameter $Q_{\tilde{K},i}$, i.e., curves with a lighter shade of  blue correspond to lower values of $Q_{\tilde{K},i}$.
In accordance with the suppression found in the primordial power spectrum, the various blue curves in Fig.~\ref{fig_Cl_Dust} show a suppression in the low multipoles that becomes stronger and affects ever high $\ell$ as the value of $Q_{\tilde{K}}$ increases. Notice that in the case of $Q_{\tilde{K},1}$, for which there is no visible effect on $P_\mathcal{R}$ in the observable range  $k\gtrsim k_0$, the angular power spectrum is almost indistinguishable from the red-dashed curve. Contrary to the cases of $n=0$ and the model analysed in Ref.~\cite{Morais2017},  we do not observe the presence of extra peaks in the angular power spectra in the low multipoles. This can be related to the absence of distinguishable peaks in the primordial power spectrum. Nevertheless, we can impose an upper bound on the free parameter by requiring that the angular spectrum $D_\ell^{TT}$ follows the Planck fit for  $\ell>30$:
\begin{align}
	\label{upperlimit_dust}
	Q_{\tilde{K}}\lesssim k_0^{3/2}\approx3.038\times10^{-4}k_*^{3/2}
	\,, 
\end{align}
which is well inside the range in Eq.~\eqref{constraint2a}.

We have also calculated the reduced $\chi^2$-values for this case according to \eqref{chi2} and listed them on the right-hand side of Table \ref{tabchi}. In the present case of a dust-like pre-inflationary phase, our model ``over-fits'' the CMB data for the values $Q_{\tilde{K},3}$ to $Q_{\tilde{K},6}$, i.\,e.~all $Q_{\tilde{K}}$ within a range of
\be \label{bestfit1}
0.075\,k_0^{3/2} \lesssim Q_{\tilde{K}} \lesssim k_0^{3/2}
\ee
have a $\chi^2$-value of less than unity and thus describe the measured CMB data equally well. The reason is, of course, that compared to the previous case there is no enhanced bump present, such that the apparent suppression of power on the largest scales in the CMB data can be well fitted by a wide range of parameters. 

This also reflects the fact that it is in general difficult to distinguish different approaches that attempt to explain the CMB quadrupole anomaly, given that the measurement errors in the CMB data are large on these scales, especially because of cosmic variance which is also most prominent on the largest scales. In our case this issue could only be overcome if we were able to restrict the range of the quantities that make up the parameter $Q_{\tilde{K}}$ further, i.e.~by applying some more fundamental theory of quantum gravity.

\subsection{Implications for the multiverse}

In the previous section we have obtained upper limits \eqref{upperlimit_radiation} and \eqref{upperlimit_dust} for the free parameter $Q_{\tilde{K}}$ by requiring that the angular power spectrum of the TT component of the CMB is not affected for multipoles above the $\ell=30$ threshold. This parameter is related to the $\tilde{K}$-number and the coupling parameter $\lambda_*$ via Eq.~\eqref{QK_def} and after substitution of \eqref{upperlimit_radiation} and \eqref{upperlimit_dust} we arrive at
\begin{align}
	\label{multiverse_limit}
	\frac{\lambda_*}{a_*}\tilde{K}
	=
	 2\pi\frac{\bar\lambda_*}{a_*}\frac{\tilde{K}}{\Nuni}
	 \lesssim
	 \left(\frac{k_0}{k_*}\right)^{2-n} k_*
	 \,,
\end{align}
where, on the first equality we have reversed the rescaling of the universe--universe coupling introduced in Sect.~\ref{section2}. Recalling that, by assumption, $\tilde{K}\ll \Nuni$, we can restate \eqref{multiverse_limit} as
\begin{align}
	\label{multiverse_limit_2}
	 \frac{\tilde{K}\bar\lambda_*}{\Nuni} \ll&~ \frac{k_*}{2\pi}\left(\frac{k_0}{k_*}\right)^{2-n}\, a_*
	\approx 
	\begin{cases}
		1.558 \times10^{4}\, \MP^{-1}\,,& \textrm{for }n=0\,,\vspace{5pt}\\
		2.318 \times10^{5}\, \MP^{-1}\,,& \textrm{for }n=1/2\,.
	\end{cases}
\end{align}
Notice the dependence of  \eqref{multiverse_limit_2} on the type of pre-inflation through the exponent $(k_*/k_0)^n$. In cases where the rate of deceleration is slower, i.e., for higher $n<2$, the upper limit on the ratio $\tilde{K}\bar\lambda_*/\Nuni$ becomes higher, allowing for stronger universe--universe couplings for the same $\tilde{K}/\Nuni$.

In Ref.~\cite{RP2016} it is argued that due to a mechanism of vacuum decay, the most probable multiverse configuration corresponds to the case of small $\tilde{K}$. If we consider the minimal case of $\tilde{K}=1$, the limit \eqref{multiverse_limit_2} can be translated into a constraint for the universe--universe coupling density $\bar\lambda_*/\Nuni$ which tells us that the maximum allowed value for $\bar\lambda_*$ is proportional to the total number of universes considered. As such, stronger couplings require a higher number of universes in order for the imprints on the primordial power spectrum and the CMB to not rule out the model. Notice, however, that if the number of universes considered is increased for a fixed $\bar\lambda_*$, the effects on $P_\mathcal{R}$ and $D_\ell^{TT}$ may be red-shifted to very large scales which are inaccessible to us.

If we take a look at the preferred values for $Q_{\tilde{K}}$ according to the $\chi^2$-test, we can use the best-fit values given in \eqref{bestfit0} and \eqref{bestfit1} to determine the corresponding ratios $\tilde{K}\bar\lambda_*/\Nuni$. For the radiation-like phase with $n=0$, we obtain
\begin{align}
	\label{multiverse_bf_0}
	 \frac{\tilde{K}\bar\lambda_*}{\Nuni}\biggr|_{\text{best-fit}}^{n=0} \simeq \frac{0.316\,k_0^2a_*}{2\pi k_*}\approx 4.926 \times10^{3}\,\MP^{-1}\,,
\end{align}
while for the dust-like phase with $n=1/2$, the upper limit remains at the value given in \eqref{multiverse_limit_2}, i.e.
\begin{align}
	\label{multiverse_bf_1_max}
	 \frac{\tilde{K}\bar\lambda_*}{\Nuni}\biggr|_{\text{best-fit,max}}^{n=1/2} \lesssim \frac{k_0^{3/2}a_*}{2\pi k_*^{1/2}}\approx 2.318\times10^{5}\, \MP^{-1}\,.
\end{align}
whereas for the lower limit we get 
\begin{align}
	\label{multiverse_bf_1_min}
	 \frac{\tilde{K}\bar\lambda_*}{\Nuni}\biggr|_{\text{best-fit,min}}^{n=1/2} \gtrsim \frac{0.075\,k_0^{3/2}a_*}{2\pi k_*^{1/2}}\approx 1.738\times10^{4}\, \MP^{-1}\,.
\end{align}
Obviously, we would need to individually restrict the allowed range of the three parameters further by other means in order to be able to draw conclusions with regard to the magnitude of the inter-universal coupling strength or the number of universes that give rise to a pre-inflationary phase in agreement with the CMB data. However, one of the important outcomes of the present work is that it shows that there is room already in the current data to check some of the multiverse proposals, which is by itself a major achievement.

\section{Conclusions and outlook}
\label{conclusions}

In this work, we have presented a toy model for an interacting multiverse composed of canonically quantized universes. We postulated an interaction on the level of a field theory of wave functions on superspace. A semiclassical approximation then revealed that this interaction leads to a pre-inflationary phase in the individual universes. Because of this, the scalar perturbations we analyzed within the model got suppressed on large scales and we found that this suppression can coincide with the measured quadrupole anomaly depending on a set of three free parameters, the assumed inter-universal coupling strength $\bar\lambda_*$, the number of universes in the multiverse model $\Nuni$ and the ``label'' of the sub-universe $\tilde{K}$. Unfortunately, at this point we are only able to restrict the range of these parameters with regard to the ratio $\tilde{K}\bar\lambda_*/\Nuni$ and we would need to motivate the values of a subset of these parameters using e.g.~a more fundamental theory of quantum gravity.

Nevertheless, the result that such an inter-universal interaction can lead to a prominent pre-inflationary phase looks encouraging and should motivate further studies.

However, we also have to state that our model is built on assumptions that still need to be further robustified. It is, for example, not clear on what kind of (Hilbert) space the inter-universal interaction is constructed and especially the scale-factor dependence of the inter-universal interaction term was introduced ad hoc such that a more fundamental theory is needed to justify the specific choice for the interaction.

As for work going further, it would be interesting to use such a more fundamental theory or more physically rich models to motivate a different interaction scheme that leads to a more unique pre-inflationary phase, which cannot be mimicked by the standard matter types. This is especially important due to the fact that the model choices we made for the function describing the interaction led to results that ``overfit'' the Planck data, i.\,e.~allow for a wide range of parameters to suitably describe the CMB data, specifically because of the large observational error present on large scales in particular due to cosmic variance.

Additionally, it might be fruitful to take the semiclassical approximation of the Wheeler--DeWitt equation with the inter-universal interaction term further and thus to calculate the magnitude of quantum-gravitational effects in the pre-inflationary phase, which are supposedly larger than quantum gravity effects in a universe undergoing standard inflation.

All in all, we have shown that our toy model of an interacting quantum multiverse leads to interesting effects regarding a pre-inflationary phase that can modify the power spectra significantly on large scales, which should be the basis for further studies especially with regard to finding a more robust footing for the model and to further restrict the range of the free parameters, such that eventually more conclusions can be drawn with regard to the nature and properties of the interacting multiverse in such models.

\section*{Acknowledgements}
This article is based upon work from COST Action CA15117 ``Cosmology and Astrophysics Network for Theoretical Advances and Training Actions (CANTATA)'', supported by COST (European Cooperation in Science and Technology). The research of M.\,B.-L.~is supported by the Basque Foundation of Science Ikerbasque. She and J.\,M.~also would like to acknowledge the partial support from the Basque government Grant No.~IT956-16 (Spain) and the project FIS2017-85076-P (MINECO/AEI/FEDER, UE). The research of M.\,K.~was financed by the Polish National Science Center Grant No.~DEC-2012/06/A/ST2/00395 and by the European Research Council Grant No.~ERC-2013-CoG 616732 HoloQosmos. M.\,K.~would like to thank the \emph{Department of Theoretical Physics and History of Science} of the University of the Basque Country (UPV/EHU) for its kind hospitality while part of this work was done. He also acknowledges financial support from the Basque government Grant No.~IT979-16 and from the above-mentioned COST Action CANTATA for his visits to the UPV/EHU. M.\,K.~furthermore thanks M.\,P.~D\k{a}browski for fruitful discussions. J.\,M.~would like to thank UPV/EHU for a PhD fellowship. The authors also thank J.~J.~Blanco-Pillado and D.~Brizuela for valuable discussions and comments.

\appendix
\section{The semiclassical approximation} \label{app_semi}
We follow \cite{hh85,ck87} and use the following product ansatz for the wave function:
\be \label{psisplit}
\Psi_{\tilde{K}}
\big(\alpha,\{v_k\}\big) =
\psi_{0,\tilde{K}}
(\alpha)\prod_{k}\psi_{k,\tilde{K}}
(\alpha,v_k)\,.
\ee
Inserting this ansatz into \eqref{WDW_Ktilde2} and dividing by $\Psi_{\tilde{K}}$ leads to:
\begin{align}
	\frac{\E^{-2\alpha}}{m_\text{P}^2}&\left(
		\frac{1}{\psi_{0,\tilde{K}}}\del{^2\psi_{0,\tilde{K}}}{\alpha^2}
		+\frac{2}{\psi_{0,\tilde{K}}}\del{\psi_{0,\tilde{K}}}{\alpha}\sum_k\frac{1}{\psi_{k,\tilde{K}}}\del{\psi_{k,\tilde{K}}}{\alpha}
		+\sum_k\frac{1}{\psi_{k,\tilde{K}}}\del{^2\psi_{k,\tilde{K}}}{\alpha^2}
	\right.
	\nonumber\\
	&~
	\left.
	+\sum_{k \neq k'}\frac{1}{\psi_{k,\tilde{K}}\psi_{k',\tilde{K}}}\del{\psi_{k,\tilde{K}}}{\alpha}\del{\psi_{k',\tilde{K}}}{\alpha}
	\right)  
	+m_\text{P}^2\mathcal{V}_{\tilde{K}}(\alpha) 
	+\frac{1}{\Psi_{\tilde{K}}} \sum_k\mathcal{H}_{k,\tilde{K}}\Psi_{\tilde{K}}
	= 0\,.
\end{align}
The assumption \eqref{psisplit} only holds if we can introduce a function $F(\alpha)$, such that we can split the equation into two parts (separation of variables). The first equation then reads:
\be
\frac{\E^{-2\alpha}}{m_\text{P}^2}\,\del{^2\psi_{0,\tilde{K}}}{\alpha^2} + m_\text{P}^2\left(\E^{4\alpha}\HdS^2 + \lambda(\alpha)^2 \tilde{K}^2\right)\psi_{0,\tilde{K}} = -2F(\alpha)\,\psi_{0,\tilde{K}}\,.
\ee
We can in particular set $F(\alpha) \equiv 0$, if we assume that there is no backreaction from the perturbations, and use the WKB ansatz with the function $S_0(\alpha)$ for the background wave function $\psi_{0,\tilde{K}}$
\be \label{WKBans}
\psi_0 = \E^{\I\,m_\text{P}^2\,S_{0,\tilde{K}}
(\alpha)} \,.
\ee
This leads to the Hamilton--Jacobi equation:
\be \label{HJeq}
\left(\frac{\partial S_{0,\tilde{K}}}{\partial\alpha}\right)^2 
- \E^{6\alpha}H_0^2 -  \E^{2\alpha}\lambda(a)^2 \tilde{K}^2 =0\,.
\ee
The second equation involving also the perturbation wave functions  $\psi_{k,\tilde{K}}$ is thus given by:
\begin{align} \label{split2}
\frac{\E^{-2\alpha}}{m_\text{P}^2}&\left(\frac{2}{\psi_{0,\tilde{K}}}\del{\psi_{0,\tilde{K}}}{\alpha}\sum_k\frac{1}{\psi_{k,\tilde{K}}}\del{\psi_{k,\tilde{K}}}{\alpha}+\sum_k\frac{1}{\psi_{k,\tilde{K}}}\del{^2\psi_{k,\tilde{K}}}{\alpha^2}+\sum_{k \neq k'}\frac{1}{\psi_{k,\tilde{K}}\psi_{k',\tilde{K}}}\del{\psi_{k,\tilde{K}}}{\alpha}\del{\psi_{k',\tilde{K}}}{\alpha}\right) \nonumber\\
&+\frac{2}{\Psi_{\tilde{K}}}\sum_k\mathcal{H}_{k,\tilde{K}}\Psi_{\tilde{K}}
 = 2F(\alpha) \,.
\end{align}
If we assume that
\be \label{bg-gg-pert}
\left|\frac{1}{\psi_{0,\tilde{K}}}\del{\psi_{0,\tilde{K}}}{\alpha}\right| \gg \left|\frac{1}{\psi_{k,\tilde{K}}}\del{\psi_{k,\tilde{K}}}{\alpha}\right|
\ee
and using a random phase approximation in the sense that the terms $\frac{1}{\psi_{k,\tilde{K}}}\del{\psi_{k,\tilde{K}}}{\alpha}$ add up incoherently in the sum of the third term, we can neglect this term. Furthermore splitting up $F(\alpha)$ as
\be
F(\alpha) = \sum_k f_k(\alpha)
\ee
allows us to dispose of the sum over $k$ and by multiplying with $\psi_{k,\tilde{K}}$ we get:
\be
\frac{\E^{-2\alpha}}{m_\text{P}^2}\left(\frac{2}{\psi_{0,\tilde{K}}}\del{\psi_{0,\tilde{K}}}{\alpha}\del{\psi_{k,\tilde{K}}}{\alpha}+\del{^2\psi_{k,\tilde{K}}}{\alpha^2}\right)+2\left(\frac{\mathcal{H}_{k,\tilde{K}}\Psi_{\tilde{K}}}{\Psi_{\tilde{K}}}\right)\psi_{k,\tilde{K}} = f_k(\alpha)\psi_{k,\tilde{K}}\,.
\ee
We now also neglect the second term in the first brackets based on the assumption that the wave functions $\psi_{k,\tilde{K}}$ vary much less with the background variable $\alpha$ than $\psi_{0,\tilde{K}}$ and using again \eqref{bg-gg-pert} together with a random phase approximation we can approximate the last term on the left-hand side as
\be
\left(\frac{\mathcal{H}_{k,\tilde{K}}\Psi_{\tilde{K}}}{\Psi_{\tilde{K}}}\right)\psi_{k,\tilde{K}} \approx \mathcal{H}_{k,\tilde{K}}\psi_{k,\tilde{K}}\,.
\ee
Given that by assuming that no backreaction is present we can again set all of the $f_k(\alpha) = 0$, we arrive at:
\be
\frac{\E^{-2\alpha}}{m_\text{P}^2}\frac{1}{\psi_{0,\tilde{K}}}\del{\psi_{0,\tilde{K}}}{\alpha}\del{\psi_{k,\tilde{K}}}{\alpha}+\mathcal{H}_{k,\tilde{K}}\psi_{k,\tilde{K}} = 0\,.
\ee
Using the WKB ansatz \eqref{WKBans}, we can write:
\be
\I\,\E^{-2\alpha}\del{S_{0,\tilde{K}}}{\alpha}\del{\psi_{k,\tilde{K}}}{\alpha}+\mathcal{H}_{k,\tilde{K}}\psi_{k,\tilde{K}} = 0\,.
\ee
This allows us to define the so-called WKB conformal time
\be
\del{}{\eta} := -\E^{-2\alpha}\del{S_{0,\tilde{K}}}{\alpha}\del{}{\alpha}\,,
\ee
such that we can finally express \eqref{split2} in terms of a Schr\"odinger equation for each perturbation mode:
\be \label{Schr_vk}
\I\,\del{}{\eta}\,\psi_{k,\tilde{K}} = \mathcal{H}_{k,\tilde{K}}\psi_{k,\tilde{K}} = \frac{1}{2}\left(-\,\del{^2}{v_{k}^2} + \omega^2_{k,\tilde{K}}(\eta)\,v_{k}^2\right)\psi_{k,\tilde{K}}\,.
\ee


\end{document}